\documentclass[a4paper,11pt]{article}
\usepackage[utf8]{inputenc}
\pdfoutput=1

\usepackage{jheppub}
\usepackage{subfig}
\usepackage{tikz}
\usepackage{multirow}
\allowdisplaybreaks

\definecolor{lime}{HTML}{A6CE39}
\DeclareRobustCommand{\orcidicon}{\hspace{-1mm}
	\begin{tikzpicture}
	\draw[lime, fill=lime] (0,0) 
	circle [radius=0.16] 
	node[white] {{\fontfamily{qag}\selectfont \tiny \,ID}};
	\draw[white, fill=white] (-0.0525,0.095) 
	circle [radius=0.007];
	\end{tikzpicture}
	\hspace{-3mm}
}
\foreach \x in {A, ..., Z}{\expandafter\xdef\csname orcid\x\endcsname{\noexpand\href{https://orcid.org/\csname orcidauthor\x\endcsname}
			{\noexpand\orcidicon}}
}

\title{\color{black}{\bf Dark matter substructures affect dark matter-electron scattering in xenon-based direct detection experiments}}

\author[a]{Tarak Nath Maity\orcidA{},} 
\author[a]{Ranjan Laha\orcidB{}}

\affiliation[a]{Centre for High Energy Physics, Indian Institute of Science, C.\,V.\,Raman Avenue, Bengaluru 560012, India}

\date{}
\emailAdd{tarak.maity.physics@gmail.com}
\emailAdd{ranjanlaha@iisc.ac.in}

\abstract{
Recent sky surveys have discovered a large number of stellar substructures. It is highly likely that there are dark matter (DM) counterparts to these stellar substructures. We examine the implications of DM substructures for electron recoil (ER) direct detection (DD) rates in dual phase xenon experiments. We have utilized the results of the LAMOST survey and considered a few benchmark substructures in our analysis. Assuming that these substructures constitute $\sim 10\%$ of the local DM density, we study the discovery limits of DM-electron scattering cross sections considering one kg-year exposure and 1, 2, and 3 electron thresholds. With this exposure and threshold, it is possible to observe the effect of the considered DM substructure for the currently allowed parameter space. We also explore the sensitivity of these experiments in resolving the DM substructure fraction. For all the considered cases, we observe that DM having mass $\mathcal{O}(10)\,$MeV has a better prospect in resolving substructure fraction as compared to $\mathcal{O}(100)\,$MeV scale DM. We also find that within the currently allowed DM-electron scattering cross-section; these experiments can resolve the substructure fraction (provided it has a non-negligible contribution to the local DM density) with good accuracy for $\mathcal{O}(10)\,$MeV DM mass with one electron threshold.}

\begin{document}

\maketitle

\section{Introduction}
\label{sec:intro}
It is important to leave no stone unturned in the search for the DM identity. Numerous astrophysical and cosmological observations infer the irrefutable evidence of DM \cite{Bertone:2004pz,Lin:2019uvt,Slatyer:2021qgc,Planck:2018vyg}.  Despite these insurmountable evidences of the gravitational interaction of DM, we do not yet know if the DM candidate interacts via other forces. Numerous experiments have been performed to discover the non-gravitational signature of DM, but none of them have revealed a positive result. The DD experiments have been playing a pivotal role in their quest for the DM identity. The typical nuclear recoil (NR) DD experiments, searching for weak-scale DM, have made extraordinary progress\,\cite{SuperCDMS:2015eex, Akerib:2016lao, Cui:2017nnn, DarkSide:2018bpj, XMASS:2018bid, Aprile:2018dbl, EDELWEISS:2019vjv, Amare:2019jul, CRESST:2019jnq, CDEX:2019hzn, Adhikari:2018ljm, PandaX-4T:2021bab, DEAPCollaboration:2021raj, Schumann:2019eaa, DelNobile:2021wmp, Cooley:2021rws, Aalbers:2022dzr}. Typical NR DD experiments lose their sensitivity due to kinematic mismatch for an incident non-relativistic ambient sub-GeV DM (see for instance \cite{Battaglieri:2017aum, Kahn:2021ttr, Mitridate:2022tnv, Essig:2022dfa}).\footnote{Alternatively, one can boost non-relativistic light DM through scattering with energetic particles to overcome the threshold barrier, see for e.g., \cite{Bringmann:2018cvk, Ema:2018bih, Cappiello:2019qsw, An:2017ojc, Wang:2021jic, Granelli:2022ysi, Li:2022jxo, Calabrese:2022rfa, Calabrese:2021src} or by utilizing the Migdal effect\,\cite{Ibe:2017yqa, Dolan:2017xbu, Bell:2019egg, XENON:2019zpr, Essig:2019xkx, Dey:2020sai, Knapen:2020aky, Bell:2021zkr, Bell:2021ihi, Chatterjee:2022gbo, DarkSide:2022dhx} or by absorption of fermionic DM (both for NR and ER)\,\cite{Dror:2019onn, Dror:2019dib, Dror:2020czw}.}   In order to fully characterize particle DM properties, it is important to probe DM-electron coupling too. A promising strategy to search for such DM interactions is to consider its scattering with electrons of the target materials\,\cite{Dedes:2009bk, Kopp:2009et, Essig:2011nj, Graham:2012su, Essig:2012yx, Lee:2015qva, Essig:2015cda, Roberts:2016xfw, Essig:2017kqs, Emken:2019tni, Catena:2019gfa, Bloch:2020uzh, Herrera:2021puj, Bose:2021cou}. In contrast with nuclear scattering, the maximum sensitivity to DM-electron interaction is typically achieved at a lower DM mass. For e.g., assuming a xenon target and momentum independent scattering cross-section, the maximum sensitivity is achieved at  $\sim$ 30 GeV for DM-nuclear scattering and  $\sim$ 200 MeV for DM-electron scattering. 

An ambient DM of mass $\mathcal{O}(10)$ MeV will have a kinetic energy of the $\mathcal{O}(10)$ eV, which is in the ball-park of the atomic ionization energy or the band gap energy of semiconductor. This indicates that a sub-GeV DM can ionize an electron from an atomic shell or facilitate an electron's transition from the valance band to the conduction band. Many experiments like XENON \cite{XENON:2019gfn}, SuperCDMS \cite{SuperCDMS:2018mne}, DarkSide-50 \cite{DarkSide:2018ppu, DarkSide-50:2022hin}, DAMIC \cite{DAMIC:2019dcn}, EDELWEISS \cite{EDELWEISS:2020fxc}, SENSEI \cite{Crisler:2018gci,SENSEI:2020dpa}, PandaX-II \cite{PandaX-II:2021nsg}  etc.\,\,are searching for the signatures of such a phenomenon.

The boundedness of electrons in the target material makes DM-electron scattering events inelastic. The DM velocity required to have a measurable recoil is rather high, which can be found near the tail of the DM velocity distribution (assuming that it has a Maxwell-Boltzmann form). These tails are quite sensitive to the choice of the DM velocity distribution \cite{Hryczuk:2020trm, Buch:2020xyt, Radick:2020qip, Maity:2020wic}. The present DM velocity distribution depends on the galactic structure formation history. In the well-known paradigm of $\Lambda$CDM (Lambda Cold Dark Matter), bottom-up hierarchical structure formation is a generic feature \cite{10.1093/mnras/183.3.341,Freeman:2002wq, Vogelsberger:2014kha, Springel:2017tpz, Feldmann:2022qvd, Somerville_2015, Vogelsberger:2019ynw}. Larger galaxies are formed from the merger of smaller galaxies (although the merger of similar mass galaxies may also lead to a bigger galaxy \cite{Belokurov_2018, Helmi_2018}). The gravitational field of the Milky Way (MW) is non-uniform, and this non-uniformity gives rise to strong tidal forces. When smaller galaxies accrete into the MW galaxy, the gravitational force disrupts these galaxies resulting in tidal stripping of various components (including DM) of these infalling galaxies. For an ancient merger, the DM component will have time to virialize within the MW, which may lead to an isotropic, isothermal DM halo. This scenario is often referred to as the Standard Halo Model (SHM), with the Maxwell-Boltzmann distribution representing the DM distribution. However, for relatively recent mergers, there will not be sufficient time for virialization, resulting in plenty of substructures both in the stellar and in the DM component\,\cite{Ibata:1994fv, Helmi:1999ks, Ibata:2000ys, Belokurov:2006kc, Lisanti:2011as, Myeong:2017skt, myeong2018shards, Necib:2018iwb, Necib:2019zka, Yuan_2020, 2022arXiv220102404S, 2022arXiv220102405R, 2022arXiv220611248D}. The presence of such additional stellar substructures (beyond the MW stars) have been detected by different sky-surveys like Gaia \cite{Ahn_2012,Myeong:2017skt,Belokurov_2018,2018, 2020ApJ...901...48N, 2022arXiv220611248D}, SDSS \cite{Myeong:2017skt}, LAMOST\,\cite{2018ApJS..238...16L, Yan:2022arj}, etc.,\,and have also been predicted in various N-body simulations\,\cite{Diemand:2008in, Vogelsberger:2008qb, Kuhlen:2012fz, Kuhlen:2012ft, Necib:2018igl, Simpson_2019, Helmi_2020, https://doi.org/10.48550/arxiv.2208.08443, https://doi.org/10.48550/arxiv.2208.11135}.

Since these stellar substructures arise from merged galaxies, a DM counterpart must be associated with them too (because the DM is also present in the accreted galaxies before their merger). Whether DM would follow stellar distribution or not is a matter of debate. For example, the celestial part of the Sagittarius stream might not substantially overlap with the Solar neighborhood. However, the extended DM counterpart may overlap with our local position \cite{Purcell_2012}. The similarities between DM and stellar distributions in debris flow have been pointed out in Refs.\,\cite{Lisanti:2011as, Lisanti:2014dva}. The dwarf spheroidals, which give rise to the S2-stream, are believed to have similar DM and stellar shape \cite{OHare:2019qxc} before they merged with MW. Therefore the resemblance between stellar and DM substructures is not settled yet; more dedicated studies are needed to understand this. However, the presence of this DM might manifest in the local DM density and velocity distribution: this will result in a difference of the velocity distribution from the normal MB distribution with cut off at the galactic escape velocity\,\cite{Goodman:1984dc, Drukier:1986tm}. DM DD rate is strongly dependent on the local velocity distribution of DM\,\cite{Vergados:2002hc, Green:2003yh, Ling:2009eh, McCabe:2010zh, Fox:2010bz, Fox:2010bu, Catena:2011kv, Peter:2011eu, Frandsen:2011gi, Green:2011bv, Gondolo:2012rs, DelNobile:2013cta, Mao:2013nda, Bozorgnia:2013pua, Fox:2014kua, Feldstein:2014gza, Bozorgnia:2016ogo, Gelmini:2016pei, Laha:2016iom, Benito:2016kyp, Gelmini:2017aqe, Ibarra:2017mzt, Wu:2019nhd, Bozorgnia:2017brl, Fowlie:2017ufs, Ibarra:2018yxq, Herrero-Garcia:2019ntx, Bozorgnia:2019mjk, Poole-McKenzie:2020dbo, Lawrence:2022niq}, and a different DM velocity distribution can result in a large change in our theoretical expectations. The effects of these substructures have been extensively studied in the literature in the context of typical NR DD experiments\,\cite{Gelmini:2000dm, Stiff:2001dq, Freese:2003tt, Freese:2003na, Bernabei:2006ya, Savage:2006qr, Peter:2013aha, OHare:2017rag, OHare:2018trr, Evans:2018bqy, Buckley:2019skk, Ibarra:2019jac, OHare:2019qxc, Buch:2019aiw, DEAP:2020iwi}. This paper aims to study the effect of these DM substructures in the ER DM DD experiments assuming xenon-based detectors. Such a study has been conducted for semiconductor target material in Ref.\,\cite{Buch:2020xyt}. It was shown in Ref.\,\cite{Maity:2020wic} that the effect of such astrophysical uncertainties is quite prominent for xenon targets. Further, in large regions of the DM parameter space, the sensitivity of xenon targets is a few orders of magnitude stronger than those from semiconductor-based experiments\,\cite{XENON:2019gfn, Crisler:2018gci, SENSEI:2020dpa, PandaX-II:2021nsg} implying that xenon detectors will probably play a big role in discovering DM-electron scattering. These facts motivate our detailed study in this manuscript, where we highlight the importance of considering DM substructures while searching for DM-electron scattering.

It has been argued in Refs.\,\cite{Ahn_2012, Myeong:2017skt,Belokurov_2018,2018,Necib:2018iwb, Necib:2019zka, Yuan_2020, Ou:2022wvr} that there are plenty of stellar substructures in the local halo. We utilize the results of the LAMOST survey \cite{2018ApJS..238...16L} to present the effect of the DM substructure \cite{Yuan_2020} in DM ER experiments. Without a loss of generality, we demonstrate our results by choosing a few benchmark substructures. We expect broadly similar results for other relevant substructures. In addition, our formalism will be useful for future analysis of DM ER experiments for xenon-based targets. Currently we do not understand how much of these substructures contributes to the local DM density. We have chosen a few benchmark values of DM substructure contributions to the local DM density, namely $100\%$, $20\%$, and  $10\%$ and presented our results. Our choices are motivated by Ref.\,\cite{2022arXiv220102405R} which states that stellar substructures near the Sun may constitute $\gtrsim 20\%$ of the stellar halo.  We also consider the forecast of xenon targets in resolving the fraction of DM substructures components for a few benchmark choices of the DM parameter space.

The rest of the paper is organized as follows. In Sec.\,\ref{sec:DMe}, we briefly review the DM-electron scattering in xenon-based detectors. In Sec.\,\ref{sec:DMSS}, we describe DM substructures that we have considered in our analysis. In Sec.\,\ref{sec:DMeSS}, we present our results along with the statistical methodology, and conclude in Sec.\,\ref{sec:conclusion}.        

\section{DM-electron scattering at xenon}
\label{sec:DMe}
If the ambient DM particle scatters off an electron of xenon, DM may transfer its kinetic energy to the electrons, leading to free electrons. For example, a non-relativistically moving ambient DM of mass $\sim 100 $ MeV will have kinetic energy  $\sim 50$ eV (in the Solar system), which is in the ballpark of the electron ionization energy of xenon.

In a two-phase xenon time projection chamber, DM particles interact with the liquid Xe target material, and depending on interaction type (electronic or nuclear), the signal topologies are different. For DM-nuclear interaction, the deposited DM energy produces excited atoms, electron-ion pairs, and some non-observable heat. Some free electrons recombine with ionized atoms to generate more excited atoms. Essentially both the direct and excited states produced by electron-ion recombination make a characteristic scintillation light. This prompt scintillation light, known as S1, is detected in photomultiplier tubes (PMTs) immersed in the liquid Xe at the bottom. Due to an external electric field, the remaining electrons drift through liquid xenon and cross the liquid and gaseous interface, producing proportional scintillation in the upper PMTs. This signal is known as S2. For the ER interactions, almost all the ionized electrons are collected at the upper PMTs through scintillation, producing a dominant S2 signal with a subdominant S1 signal. Hence ER interactions manifest through a large S2/S1 ratio compared to the NR case\,\cite{DiGangion:2021thw}.

Let us consider a DM particle of mass $m_{\chi}$ and velocity $v$ scattering off an electron in the xenon atom.  Energy conservation implies\,\cite{Bloch:2020uzh}
\begin{equation}
\label{eq:vmin}
v_{\rm min}=\frac{q}{2 m_{\chi}}+\frac{\Delta E_e}{q},
\end{equation}
where $v_{\rm min}$ is the minimum DM velocity required to get an ER of $\Delta E_e$, and $q$ is the momentum transfer to the electron. Note that $\Delta E_e$ must be greater than the ionization energy of the corresponding shell $E_{n,l}$ to have an observable recoil $E_e$, i.e., $\Delta E_e = E_{n,l} + E_e$. The differential DM-electron scattering event rate can be written as \cite{Essig:2017kqs}
\begin{equation}
\label{eq:rateXe}
\frac{dR}{d\,{\rm ln}\, E_e}=N_T\frac{\rho_{\chi}}{m_{\chi}}\,\sum_{nl} \frac{\bar{\sigma}_e}{8\mu_{\chi e}^2} \int q dq \,F_{\rm DM}(q)^2\, |f_{\rm ion}^{n,l}(k^{\prime},q)|^2 \,\eta\left(v_{\rm min}(k^{\prime},q),t\right),
\end{equation}
where $N_T$ is the number of electrons in the target, $\rho_{\chi}$ denotes the local DM density, and DM-electron reduced mass is represented by $\mu_{\chi e}$. DM-electron scattering cross section for a reference momentum transfer, namely $q=\alpha m_e$, is indicated by $\bar{\sigma}_e$. The DM form factor, $F_{\rm DM}(q)$, takes care of the momentum dependency in the cross-section. The ionization form factor is represented by $f_{\rm ion}^{n,l}$ with $n$ and $l$ being the principal and angular momentum quantum number, respectively. The recoil momentum is denoted by $k^{\prime}=\sqrt{2 m_e E_e}$. The time dependency of the recoil signal is described through $t$. The quantity $\eta$, also called the mean inverse speed, depends on the $i^{\rm th}$ DM velocity distribution as
\begin{equation}
\label{eq:eta}
\eta^i(v_{\rm min},t)=\int_{v_{\rm min}}^{\infty} \frac{f_{\rm lab}^i(\mathbf{v},t)}{v} d^3v,
\end{equation}
where $f_{\rm lab}^i$ is the DM velocity distribution at the detector's rest frame in the location of the Earth for the $i^{\rm th}$ DM component (which contributes to the DM velocity distribution). The latter can be obtained by boosting the galactic rest frame DM velocity distribution ($f_{\rm gal}$)   
\begin{equation}
\label{eq:galtolab}
f_{\rm lab}^i(\mathbf{v},t) = f_{\rm gal}^i(\mathbf{v+v}_{\rm E}(t)),
\end{equation}
where $\mathbf{v}_{\rm E}$ is the Earth's velocity in the galactic rest frame:
\begin{equation}
\mathbf{v}_{\rm E}(t)=\mathbf{v}_{\rm LSR}+\mathbf{v}_{\rm pec}+\mathbf{u}_{\rm E}(t).
\end{equation}
Here $\mathbf{v}_{\rm LSR}$ is the velocity of the local standard of rest (LSR), $\mathbf{v}_{\rm pec}$ is the peculiar velocity of the Sun with respect to the LSR. Conventionally these are expressed in galactic rectangular co-ordinate and expressed as $\mathbf{v}_{\rm LSR}=(0,v_0,0)$, $\mathbf{v}_{\rm pec}=(11.1 \pm 1.5, 12.2 \pm 2, 7.3 \pm 1)$ km/s \cite{Sch_nrich_2010}. Following Refs.\,\cite{Evans:2018bqy, Maity:2020wic},  throughout the paper we fix $v_0=233$ km/s. The uncertainties associated with $v_0$ and other astrophysical parameters have been studied in Refs. \cite{Hryczuk:2020trm, Radick:2020qip, Maity:2020wic} in the context of ER (see Ref.\,\cite{Chen:2021qao} for halo independent analysis). The time-dependent Earth's velocity is represented by  $\mathbf{u}_{\rm E}(t)$ which leads to the well-known annual modulation of the signal. The expression for  $\mathbf{u}_{\rm E}(t)$ can be found in \cite{McCabe:2013kea}.

The differential event rate given in Eq.\,\eqref{eq:rateXe} can be divided into three parts. The particle physics input is indicated by $\bar{\sigma}_{e}$ and $F_{\rm DM}$. Throughout our analysis, we will do a model-independent analysis with two choices of $F_{\rm DM}$: 1 and $1/q^2$, which appears in large classes of particle physics model \cite{ Holdom:1985ag,  Borodatchenkova:2005ct, Chu:2011be, Lin:2011gj, Izaguirre:2015yja, Alexander:2016aln, Boehm:2020wbt, 10.21468/SciPostPhysLectNotes.43}. We will present the results of  $F_{\rm DM}= 1$ in the main text and that of $F_{\rm DM}= 1/q^2$ in the appendix. The atomic physics part symbolized by $f_{\rm ion}^{\rm n,l}$ signify ionization probability. The numerical values of the $f_{\rm ion}^{n,l}$ is adopted from {\tt QEdark} \cite{Essig:2015cda, Essig:2017kqs, QEdark}. The local DM density and $\eta$ constitute the astrophysical inputs.

The galactic DM velocity distribution is traditionally assumed to be a Maxwell-Boltzmann (MB) distribution truncated at the galactic escape velocity ($v_{\rm esc}$)
\begin{equation}
    f_{\rm gal}^{\rm MB}(\mathbf{v})=
      \frac{1}{(2 \pi \sigma_v^2)^{3/2} N_{\rm esc}^{\rm MB}}\exp{\left(-\frac{|\mathbf{v}|^{2}}{2 \sigma_{v}^{2}}\right)} \Theta(v_{\rm esc}-|\mathbf{v}|) \,.
\label{eq:fvSHM}
\end{equation}
The isotropic velocity dispersion $\sigma_{v}$ is related to $v_0$: $v_0=\sqrt{2} \sigma_v$. The normalization constant $N_{\rm esc}^{\rm MB}={\rm erf}(z)- 2\pi^{-1/2}z e^{-z^2}$ with $z=v_{\rm esc}/v_0$ and erf is the error function. Throughout the discussion the galactic escape velocity ($v_{\rm esc}$) has been fixed to $528$\,km/s \cite{Evans:2018bqy, Deason_2019}. While the MB distribution may describe the DM velocity distribution which is in equilibrium (hydrodynamical simulations indicate that MB distributions may not adequately describe the velocity distribution of the smooth DM halo component), the equilibration condition will not be met for relatively recent mergers of the MW with other galaxies. These recent mergers will have unique signatures, both in velocity and position space, called substructures. The existence of these substructures is also observed in various N-body simulations.  When a galaxy accretes into the Milky Way, the stellar component of the accreted galaxy carries several tell-tale signatures: stellar streams, stellar shards, and stellar debris flow\,\cite{Ibata:1994fv, Helmi:1999ks, Ibata:2000ys, Belokurov:2006kc, Lisanti:2011as, Myeong:2017skt, myeong2018shards, Necib:2018iwb, Necib:2019zka, Yuan_2020, 2022arXiv220102405R, 2022arXiv220102404S, 2022arXiv220611248D}.

The recent results of various surveys like  Gaia, SDSS, and LAMOST  indeed indicate the presence of these stellar substructures. Combining the effect of the substructure with the SHM, we get total average inverse speed as
\begin{equation}
\eta(v_{\rm min},t)= \int_{v_{\rm min}}^{\infty} \frac{1}{v} \left[ (1-\delta) f_{\rm lab}^{\rm MB}(\mathbf{v},t) + \delta f_{\rm lab}^{\zeta_i}(\mathbf{v},t)\right] d^3v,
\label{eq:etacombine}
\end{equation}
where $f_{\rm lab}^{\zeta_i}(\mathbf{v},t)$ refers to the substructure velocity distribution (discussed in Sec.\,\ref{sec:DMSS}) and $\delta$ represents the fractional contribution that the corresponding component constitutes to the local density of DM.\footnote{If each of the substructures contributes different fractions then instead of one $\delta$ there will be a set of such $\delta$'s. For simplicity, we have ignored the effect of multiple substructures.} In what follows, we will consider the effect of these substructures in DM velocity distribution and the ER DD rate in liquid xenon experiments. 
\section{DM substructures}
\label{sec:DMSS}

This section discusses the benchmark DM substructures that we have studied in this work. We have utilized the results of Ref. \cite{Yuan_2020} where the stellar substructure is obtained using the star catalog of LAMOST DR3 \cite{2018ApJS..238...16L}. We choose a few representative substructures to present our results. For clarity, we also mention the name of the associated dynamically tagged groups (DTG) with the relevant substructures \cite{Yuan_2020}. The details of these substructures are summarised in Table \ref{tab:subs}. We emphasize that the chosen substructures are for illustrative purposes only. Further research is required in order to understand the DM content of various substructures and whether the substructure DM profile coincides with the Solar circle. Whether the corresponding DM substructure will follow the same velocity distribution as the stellar substructure or not is currently not understood. Using Via Lactea II high-resolution N -body simulation, it has been shown that DM debris flows closely follows their stellar counterpart \cite{Lisanti:2011as, Lisanti:2014dva}. However, the same is not valid for Sagittarius stream \cite{Purcell_2012}.   Nevertheless, we will assume that the velocity distributions of the substructures follow that of the corresponding stellar components. This assumption can be confirmed or refuted by future research. However, the broad conclusion (like the change in the event rate and subsequently in the discovery limit due to DM substructures) of this study will hold.

We note that the substructures we have considered in this paper have similarities with previous considerations \cite{OHare:2019qxc, Buch:2020xyt}. For instance, the Helmi substructure is analogous to S2-substructure \cite{Helmi_2020}. The velocity properties of the Nyx substructure are somewhat similar to the prograde (Pg) stream and are expected to arise from the same Splashed Disk event \cite{Yuan_2020}.\footnote{Ref.\,\cite{2021ApJ...912L..30Z} has argued that Nyx is a part of thick disk.} Some of the considered substructures are also found in Gaia DR3 data at the Solar neighborhood \cite{2022arXiv220102404S, 2022arXiv220102405R, Ou:2022wvr}.

\begin{table}[]
\centering
\begin{tabular}{|c|c|c|c|c|c|c|}
\hline
\multicolumn{1}{|l|}{\multirow{2}{*}{Substructure}} & \multicolumn{3}{c|}{Mean velocity (km/s)} & \multicolumn{3}{c|}{Velocity dispersion  (km/s)} \\ \cline{2-7} 
\multicolumn{1}{|l|}{}                              & $\mu_R$    & $\mu_{\phi}$    & $\mu_z$    & $\sigma_R$    & $\sigma_{\phi}$   & $\sigma_z$   \\ \hline
HelmiDTG1                                           & 4.5        & 197.2           & 244.3      & 146.0         & 62.6              & 42.4         \\ \hline
HelmiDTG3                                           & 26.2       & 157.1           & -241.3     & 78.9          & 28.8              & 27.2         \\ \hline
PolarDTG11                                               & -47.9      & 21.8            & 229.2      & 75.4          & 19.2              & 21.5         \\ \hline
PgDTG2                                                  & 221.2      & 155.7           & 139.7      & 26.2          & 33.8              & 52.3         \\ \hline
Sausage                                             & 2.1        & -0.3            & -8.7       & 136.6         & 35.0              & 72.3         \\ \hline
RgDTG28                                                  & -4.0       & -106.1          & -143.2     & 115.8         & 29.3              & 30.3         \\ \hline
Sequoia                                             & -36.9      & -273.9          & -87.0      & 138.2         & 36.7              & 65.0         \\ \hline
\end{tabular}%
\caption{The details of the substructures are used in this paper. The numerical values of the mean velocities and diagonal values of the velocity dispersions are adapted from tables 2 and 3 of \cite{Yuan_2020}. The DTG from which substructures are identified has also been specified.}
\label{tab:subs}
\end{table}
The mean stellar velocities and the diagonal values of the stellar velocity dispersions are given in Table \ref{tab:subs}. In general, DM substructures will have a different velocity distribution than the virialized component (SHM), which will dramatically impact the ER distribution. The galactic velocity distribution for each of the substructures (referred to by $\zeta_i$) can be written as \cite{OHare:2019qxc, Buch:2020xyt}
\begin{figure*}[t]
\begin{center}
	\subfloat[\label{sf:lowfv}]{\includegraphics[angle=0.0,width=0.45\textwidth]{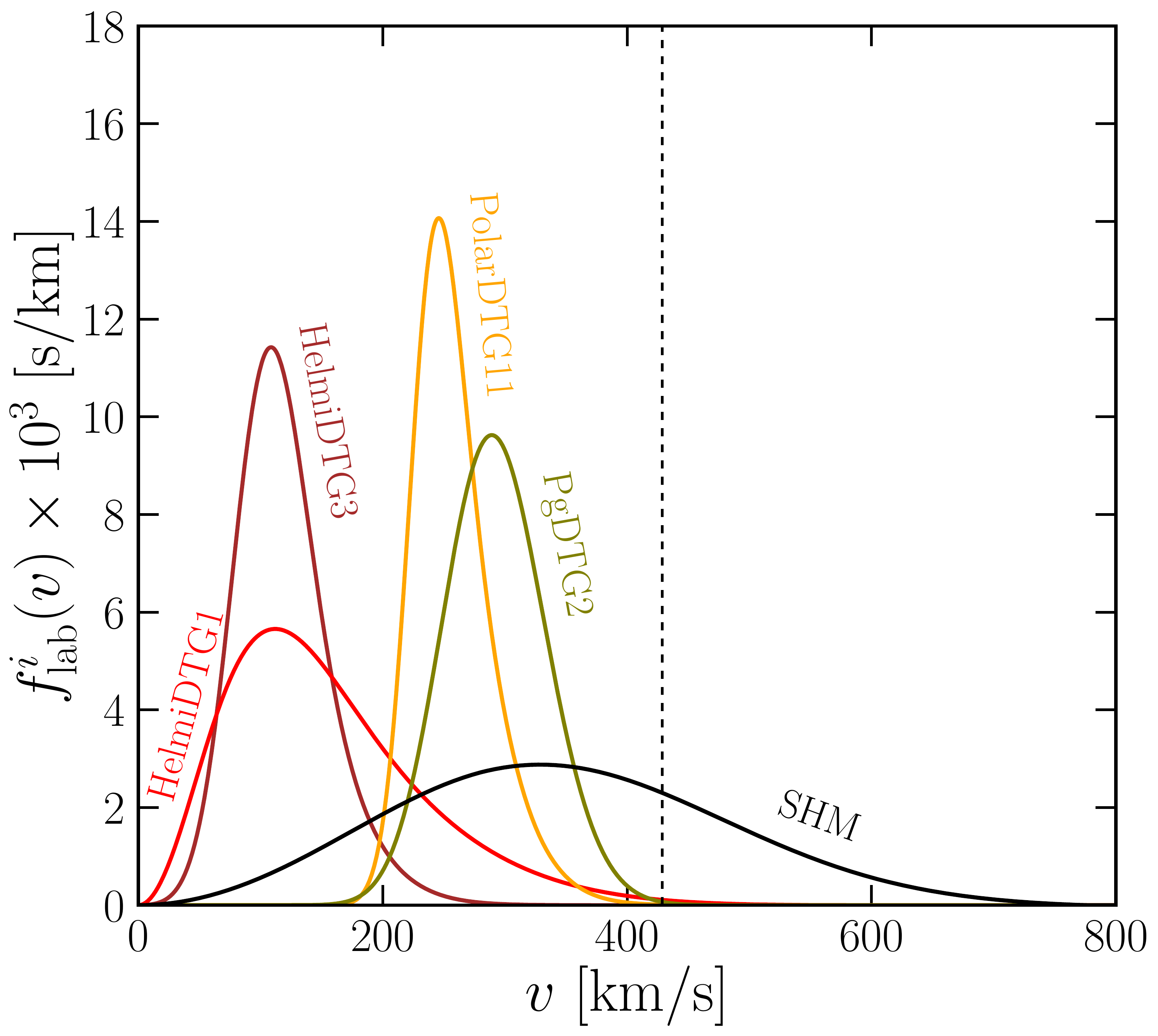}}~~
	\subfloat[\label{sf:highfv}]{\includegraphics[angle=0.0,width=0.45\textwidth]{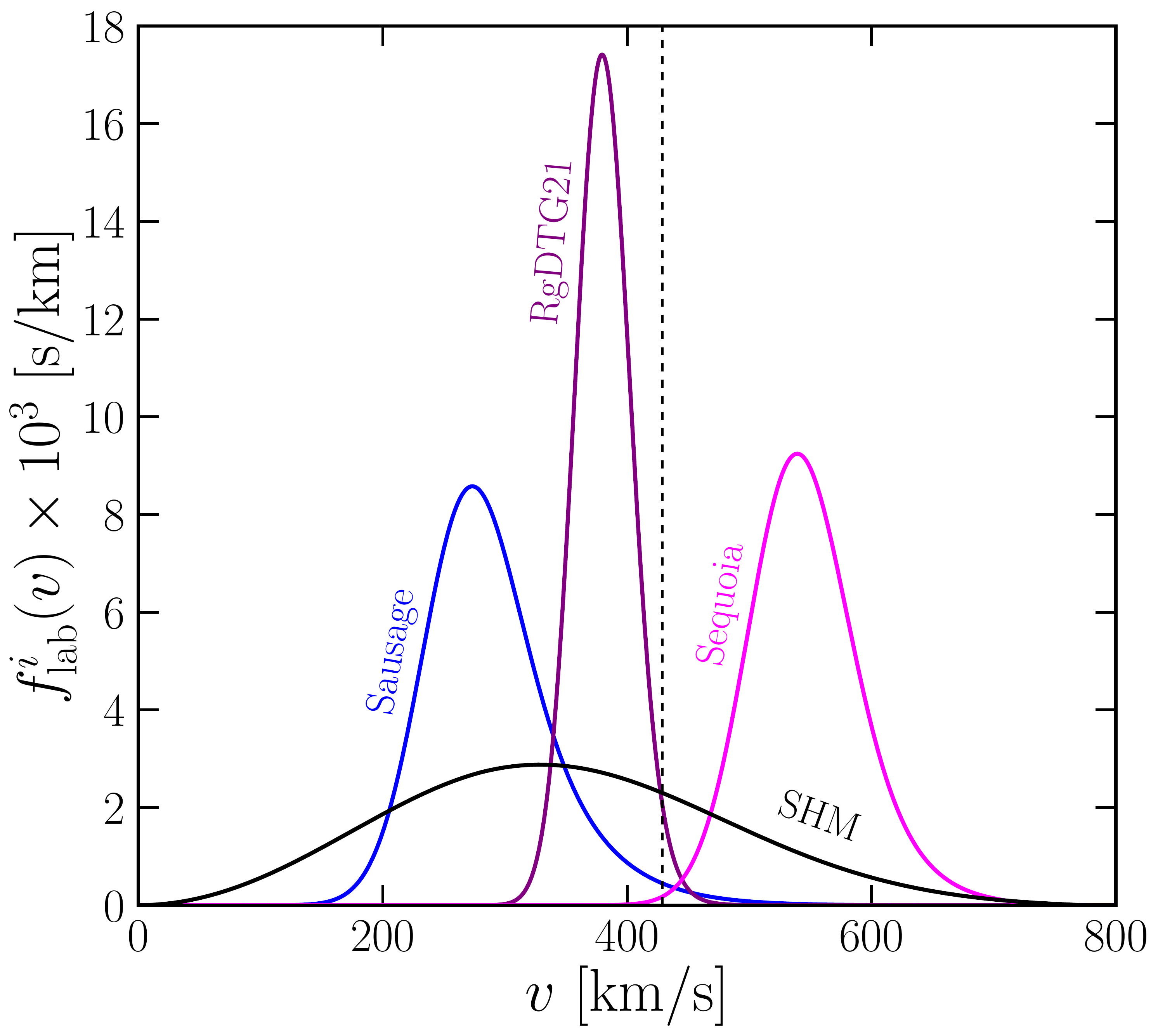}}~~\\	
	\caption{We show the lab frame DM speed distributions for all the considered astrophysical components. The red, brown, orange, olive, blue, purple, and pink solid lines represent the HelmiDTG1, HelmiDTG3, PolarDTG11, PgDTG2, Sausage, RgDTG28, and Sequoia, respectively. We also display the same for SHM by the solid black line.  The vertical dashed line shows required $v_{\rm min}$ for $m_{\chi}=100$ MeV with $E_e=20$\,eV, $q=25$\,keV, and $5p^6$ shell.}
	\label{fig:VelDist}
\end{center}	
\end{figure*}
\begin{equation}
f_{\rm gal}^{\zeta_i}(\mathbf{v})=\frac{1}{(8\pi^3 \, {\rm det}\, \sigma^{\zeta_i})^{1/2} N_{\rm esc}^{\zeta_i} } {\rm exp} \left(-(\mathbf{v}-\boldsymbol{\mu}^{\zeta_i})^T\frac{1}{2 (\sigma^{\zeta_i})^2 } (\mathbf{v}-\boldsymbol{\mu}^{\zeta_i})\right) \Theta(v_{\rm esc}-|\mathbf{v}|),
\label{eq:fvsubs}
\end{equation}
where $\sigma^{\zeta_i}$ is the velocity dispersion matrix, assumed to be diagonal with the values given in Table \ref{tab:subs} and ${\rm det}\, \sigma^{\zeta_i}$ is the determinant of the dispersion matrix.  The mean velocities of the substructures in the galactic frame are expressed by $\boldsymbol{\mu}^{\zeta_i}$ which are non-zero in contrast to the SHM case, as indicated in Table \ref{tab:subs}. The normalization constant $N_{\rm esc}^{\zeta_i}$ is calculated numerically. The step function represents the cut-off at the galactic escape velocity, although the substructures' velocity distributions are likely to peak at smaller velocities. Therefore this cut-off will have a numerically insignificant effect. The index $\zeta_i$ refers only to the substructure, whereas $i$ includes both the substructures and SHM.

Assuming Eq.\,\eqref{eq:fvsubs} as the galactic velocity distributions for the DM substructures, we display the corresponding lab frame speed distributions, $f_{\rm lab}^i(v) = v^2 \int d \Omega f_{\rm lab}^i(\mathbf{v})$, using Eq.\,\eqref{eq:galtolab} in Fig.\,\ref{fig:VelDist}. Except for the modulation signature (discussed in Sec.\,\ref{subsec:res}), we fix the Earth's velocity to $\mathbf{v_E}\,=\,(39.7, 243.2, 16.4)$\,km/s  to economize the computation. This value of $\mathbf{v_E}$ is attained during first week of March when the Earth's velocity is roughly equal to its average velocity. For Sequoia we have explicitly checked that taking the exact yearly average rate would lead to less than $\sim 6\%$ change in the discovery limit. Given the poor knowledge of DM substructure fraction, we ignored this $\lesssim 10\%$ effect.   The general trend we observe is that the substructures which peak at larger values of $v$ have negative $\mu_{\phi}$. Since the Earth moves with high positive rotational velocity  $\sim 250$ km/s, substructures with negative $\mu_{\phi}, $  will hit the Solar system with larger velocities. On the other hand, substructures having large positive $\mu_{\phi}$ co-rotate with the Earth, leading to $f_{\rm lab}^i(v)$ peaking at smaller velocities. This has been displayed in Fig.\,\ref{fig:VelDist}, where the Helmi streams having larger values of $\mu_{\phi}$ peak at relatively smaller velocities, whereas Sequoia having a negative $\mu_{\phi}$ peaks at the higher velocity. We also display the velocity distribution of SHM by the solid black line. For reference we show the required $v_{\rm min} = 428.7$ km/s to obtain a recoil of $20$\,eV with momentum transfer $25$\,keV and $5p^6$ shell for DM mass $100$\,MeV by the vertical black dashed line.

\begin{figure*}[t]
\begin{center}
	\subfloat[\label{sf:loweta}]{\includegraphics[angle=0.0,width=0.45\textwidth]{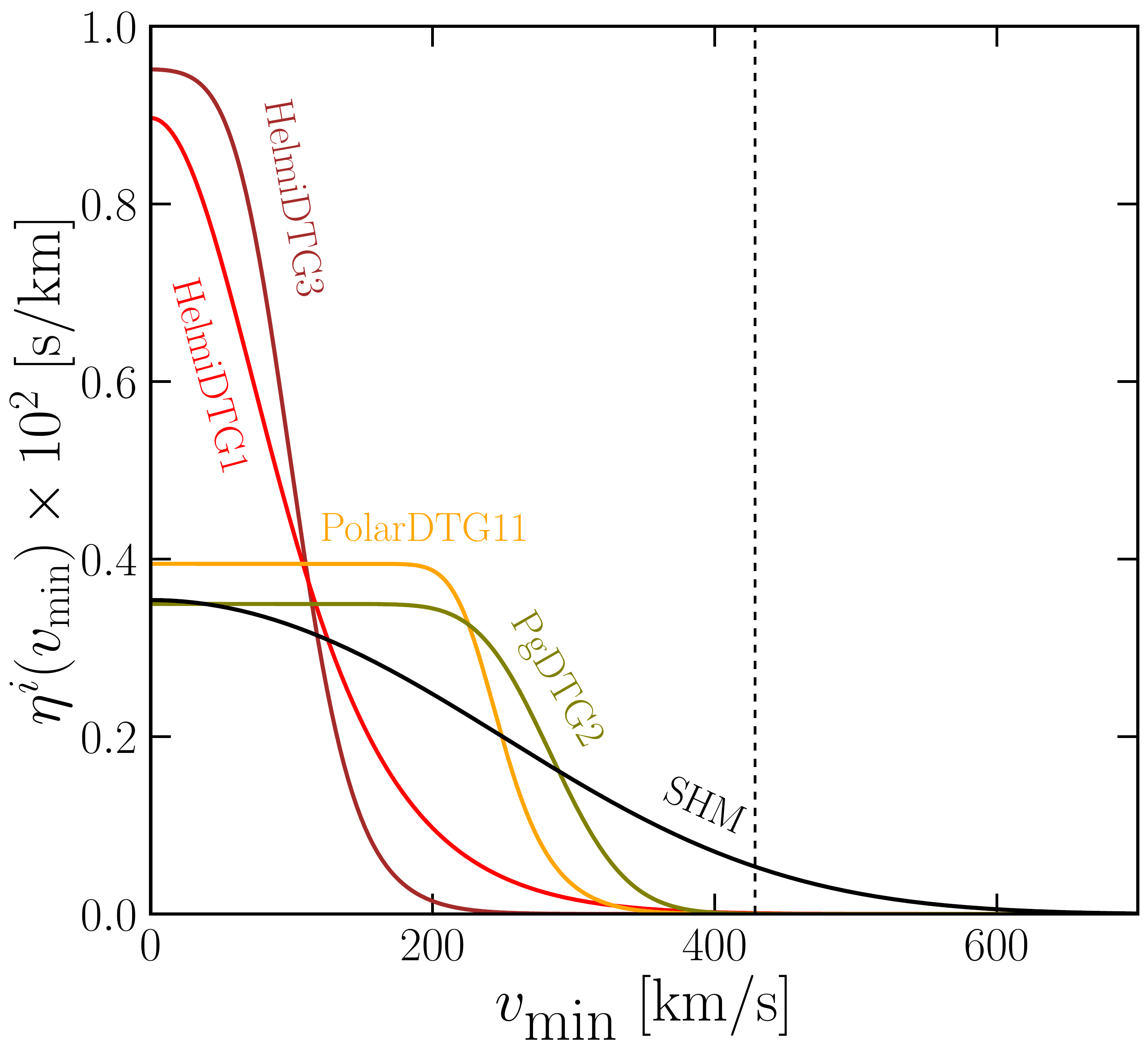}}~~
	\subfloat[\label{sf:higheta}]{\includegraphics[angle=0.0,width=0.45\textwidth]{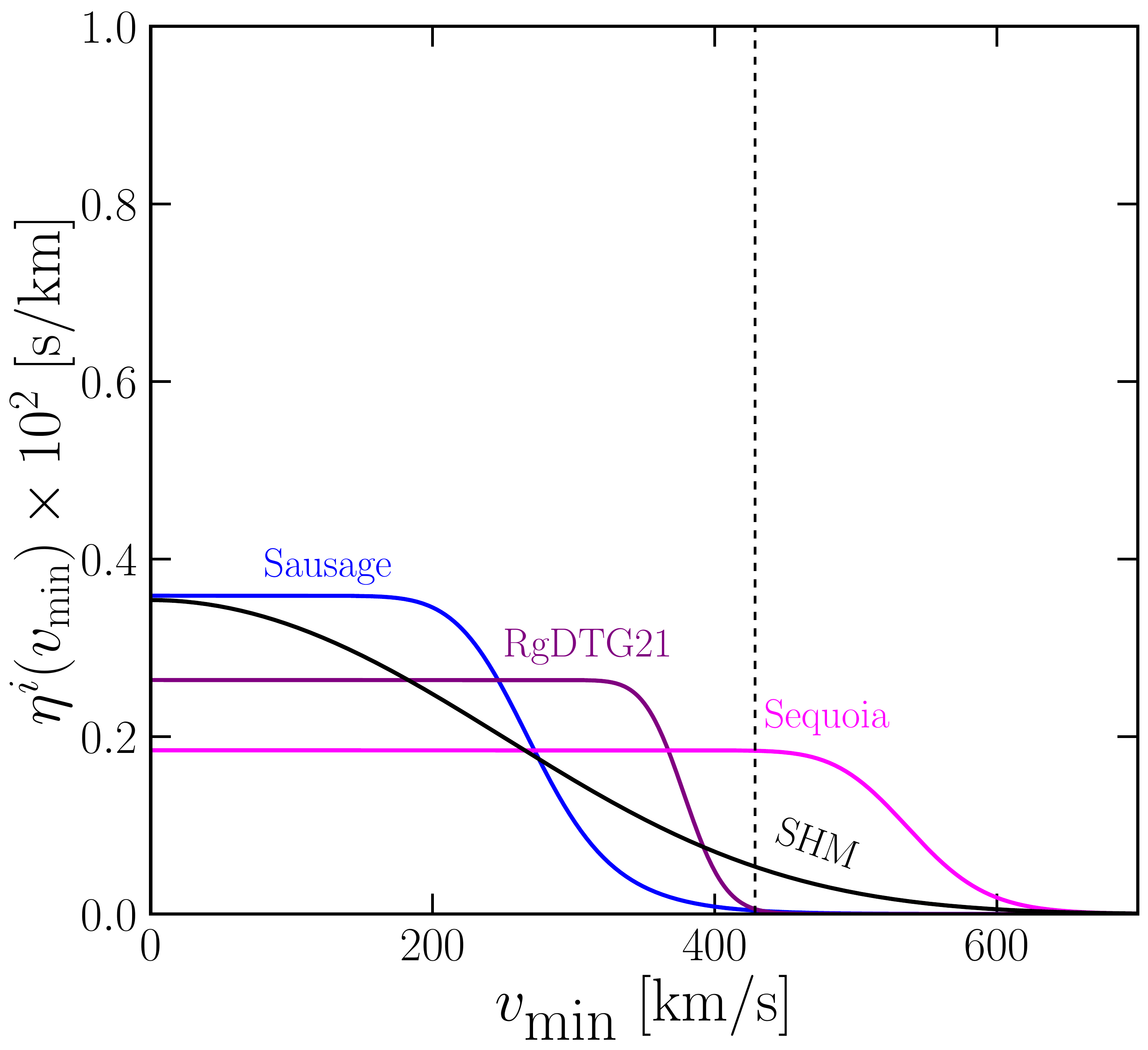}}~~\\	
	\caption{The mean inverse speed ($\eta^i(v_{\rm min})$) for the considered substructures and the SHM. The color coding and other relevant details are same as Fig.\,\ref{fig:VelDist}.}
	\label{fig:eta}
\end{center}	
\end{figure*}

Given these velocity distributions, we turn to the discussion of the mean inverse speed $\eta^i(v_{\rm min})$ (using Eq.\,\eqref{eq:eta}) of each of the astrophysical components. The values of $\eta^i(v_{\rm min})$ as a function of $v_{\rm min}$ are depicted in Fig.\,\ref{fig:eta}. The vertical black dashed line indicates values of $\eta^i(v_{\rm min})$ for $v_{\rm min} = 428.7$ km/s. Expectedly, $\eta^i(v_{\rm min})$ are monotonically decreasing function of $v_{\rm min}$, which can be understood from the integration over velocity starting from $v_{\rm min}$. The maximum values of $\eta^i(v_{\rm min})$, i.e., $\eta^i(0)$ is larger for the distributions which peak at lower velocities because the mean inverse speed is inversely proportional to the most probable speed (the speed at which velocity distribution attains maximum value) of the distribution. Hence in Fig.\,\ref{fig:eta}, we observe maximum and minimum $\eta^i(0)$ for HelmiDTG3 and Sequoia respectively. For the other distributions, $\eta^i(0)$ lie within the same of HelmiDTG3 and Sequoia. The flatness of $\eta^i(v_{\rm min})$ for Sequoia up to a large value of $v_{\rm min}$  as compared to other distributions is also a manifestation of the higher most probable speed of Sequoia. This indicates the extent to which $v_{\rm min}$ is supported by the distribution. It should also be noted that the flatness of $\eta^i(v_{\rm min})$ is also sensitive to the choice of the velocity dispersion.

%
%
%
\section{DM-electron scattering at xenon: effect of substructure}
\label{sec:DMeSS}
In this section, we discuss the effect of the substructures on the DM-electron scattering rate for liquid xenon experiments.  For $F_{\rm DM}(q)=1$, the constraint on the DM-electron scattering cross-section from the xenon detectors dominate when DM mass is $\gtrsim$ 50 MeV.  Xenon experiments may have a better prospect of discovering DM-electron scattering, and it is essential that we study this prospect thoroughly.  Our work outlines the theory effort toward answering this important question.

Following Ref.\,\cite{Essig:2017kqs}, we convert the ER energy ($E_e$) to number of electrons ($n_e$). DM-electron scattering would produce $n_e$ number of observable electrons, unobservable photons, and heat. Some primary electrons would recombine with secondary ions with probability $f_R$. Further, each recoiling electron of energy $E_e$ will give rise to additional secondary $n_e^{(1)}={\rm Floor}[E_e/W]$ quanta (photon or electron). The average energy required to create a single quanta is $W$. Moreover, the scattering process can also lead to the ionization of electrons from the inner shell, which would de-excite by releasing a photon. These photons may also create secondary quanta, $n_e^{(2)}={\rm Floor}[\Delta E_{i,j}/W]$, $\Delta E_{i,j}$ is the difference between binding energies between the relevant inner and outer shells. The number of secondary electrons produced is calculated using a binomial distribution with $n_e^{(1)}+n_e^{(2)}$ trials, having success probability $f_e$. We have chosen fiducial values (i.e., $W=13.6$\,eV, $f_e=0.83$, $f_R=0$) of the relevant parameters to  convolute Eq.\,\eqref{eq:rateXe} which will give the differential event rate as a function of number of produced electrons. Our paper does not consider uncertainties associated with $W,\,f_e,$ and $f_R$.

\begin{figure*}[t]
\begin{center}
	\subfloat[\label{sf:lowevent}]{\includegraphics[angle=0.0,width=0.45\textwidth]{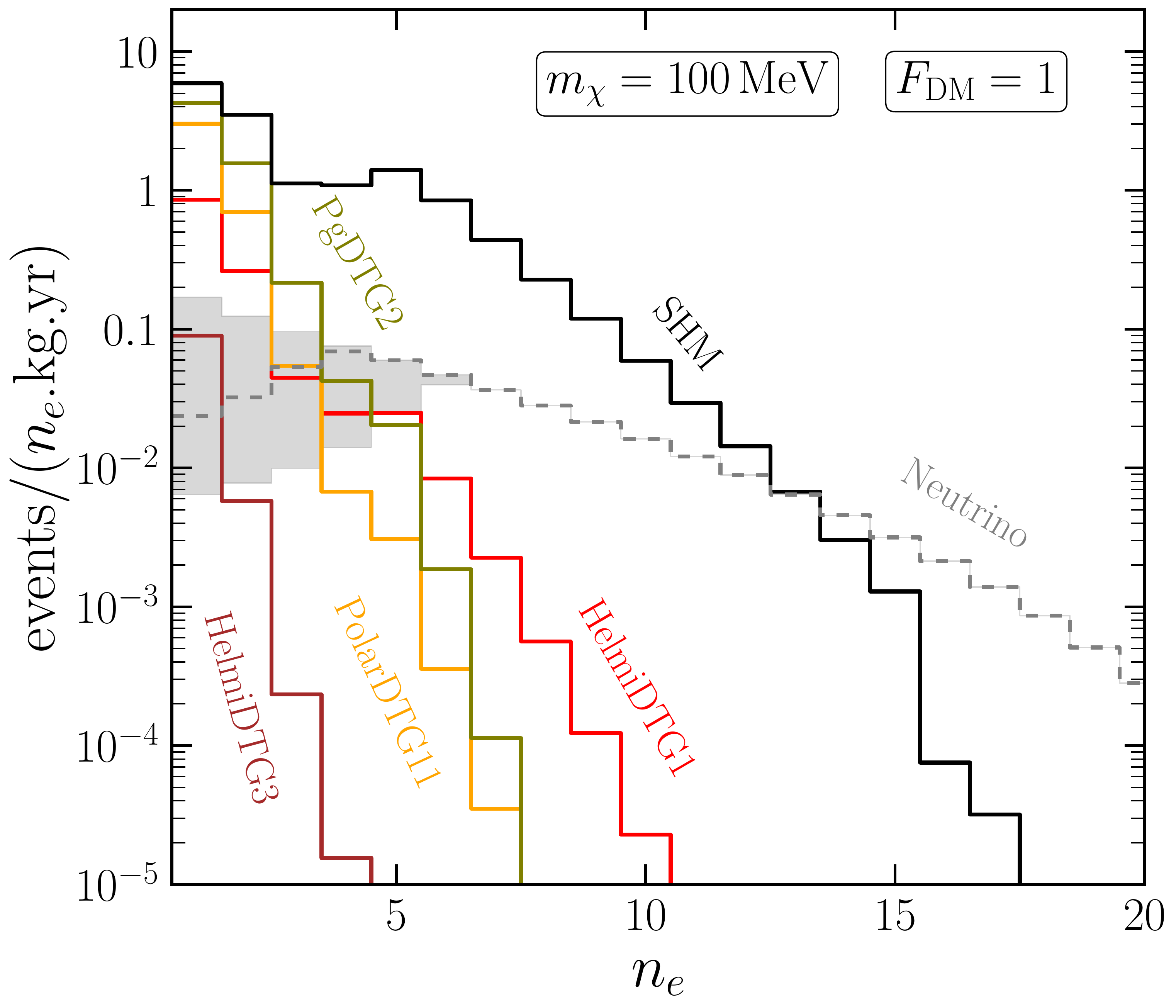}}~~
	\subfloat[\label{sf:highevent}]{\includegraphics[angle=0.0,width=0.45\textwidth]{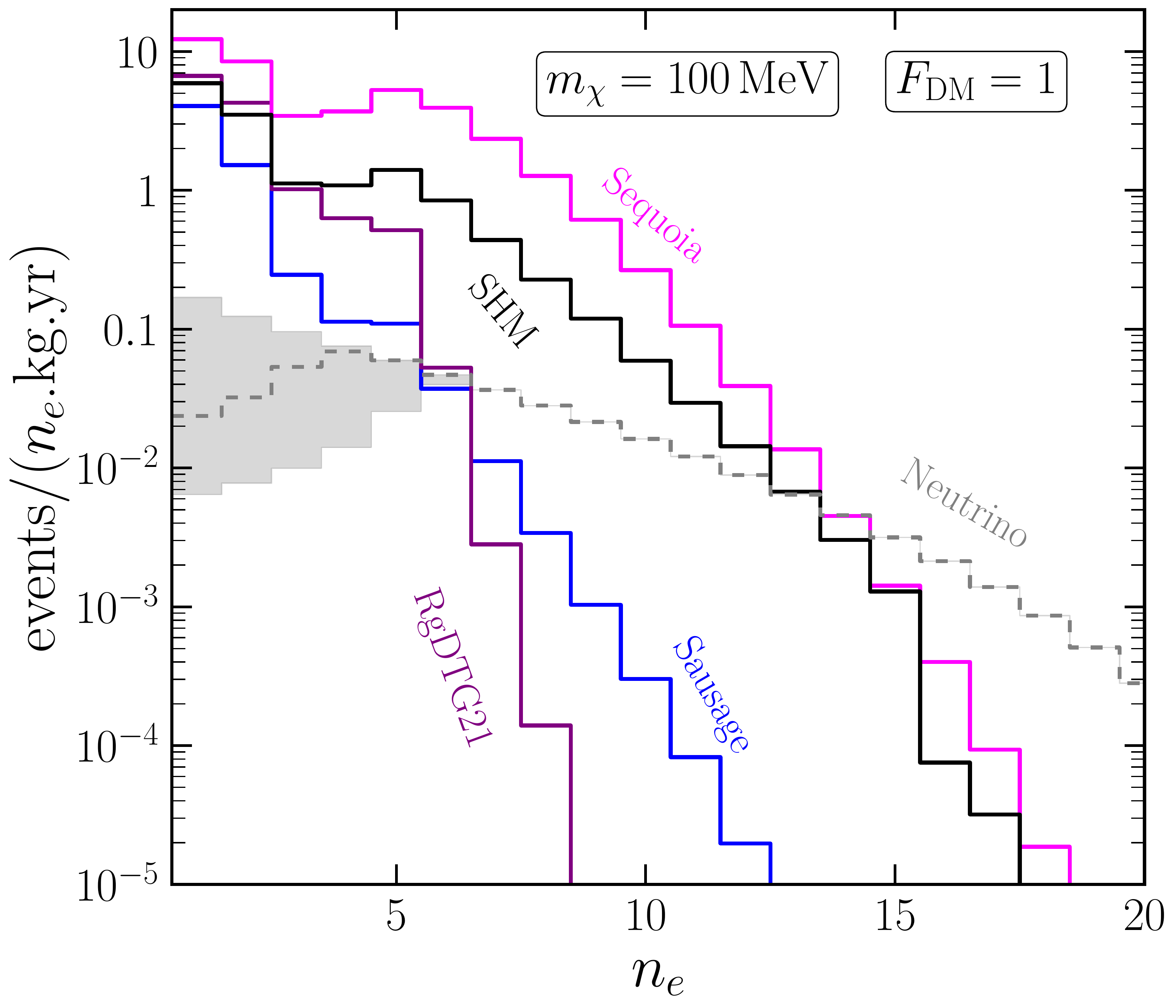}}~~\\	
	\caption{Differential event rate as a function of number of electron $n_e$ with exposure of 1 kg-year, for $m_{\chi}=100$\,MeV, and $\bar{\sigma}_e=10^{-40} \, {\rm cm}^2$ assuming each of the labelled astrophysical component contribute $100\%$ to the local DM density. The neutrino-induced event rates are illustrated by the grey dashed lines, and the grey shaded regions represent the variation in the event rates for different ionization models.}
\label{fig:EventRate}
\end{center}	
\end{figure*}

In Fig.\,\ref{fig:EventRate}, we show the differential event rate as a function of $n_e$ for $m_{\chi}=100$\,MeV, $\bar{\sigma}_e=10^{-40} \, {\rm cm}^2$, and 1 kg-year exposure. For each event rate, we have assumed that the corresponding astrophysical component (SHM or substructures) constitutes $100\%$ of the local DM density. For $m_{\chi}=100$\,MeV with typical momentum transfer of $\mathcal{O}(10)$\,keV, to obtain a measurable recoil the required minimum DM velocity should be around $500$\,km/s. Hence, the tail of $\eta^i(v_{\rm min})$ dominantly contributes to the recoil rate. Evidently the substructures having the largest value of $\eta^i(v_{\rm min})$ near $v_{\rm min}\sim 500$\,km/s give rise to a larger event rate.

\subsection{Neutrino background}
\label{subsec:nubag}
The scattering of neutrinos with electron/\,nucleon may also give rise to ionization signals in low-threshold DD experiments. Other background sources like radioactive background, Cherenkov radiation, etc.\,which can potentially mimic a DM signal \cite{Du:2020ldo}. The experimental collaborations confront and beat these non-neutrino backgrounds using various experimental techniques to isolate a potential DM signal. However, the neutrinos are an irreducible background that can not be removed by using shielding, purified detector material, and other experimental techniques. Because of this, we have taken neutrinos as the only source of background in our analysis. If other non-neutrino backgrounds are found in the data-set, then our results will degrade proportionally.

It has been argued in Refs.\,\cite{Essig:2018tss, Wyenberg:2018eyv} that Solar neutrinos are the main source of background for sub-GeV DM-electron scattering.\footnote{See Refs. \cite{Essig:2018tss, Schwemberger:2022fjl} or discussion related to the prospect of these detectors in probing beyond SM interactions of neutrino.} Neutrino-electron elastic scattering is the dominant contribution of background events for rather large recoil energy ($\sim 10^5$\,eV). Instead, coherent neutrino-nucleon scattering may produce small ionization, which would be the dominant source of background in our consideration. The neutrino-nucleon scattering event rate is \cite{Billard:2013qya, Essig:2018tss}
\begin{equation}
\frac{dR}{dE_{\rm NR}}=N_T M T \int_{E_{\nu}^{\rm min}} \frac{d\sigma}{dE_{\rm NR}} \frac{d\phi_{\nu}}{dE_{\nu}} dE_{\nu},
\label{eq:nurate}
\end{equation}
where $N_T,\, M$, and $T$ are the number of target nuclei per unit mass, total mass, and time respectively. The minimum neutrino energy to produce a nuclear recoil of energy $E_{\rm NR}$ is expressed by $E_{\nu}^{\rm min}=\sqrt{m_N E_R/2}$. The differential coherent neutrino nucleon cross section and the differential neutrino flux are denoted by $d\sigma/dE_{\rm NR}$ and  $d\phi_{\nu}/dE_{\nu}$ respectively \cite{Essig:2018tss,OHare:2016pjy}. We have utilized low, fiducial, and high ionization models given in Ref.\,\cite{Essig:2018tss} to obtain number of electron $n_e$ for a particular nuclear recoil energy. The corresponding neutrino-induced event rate for fiducial model is displayed in Fig.\,\ref{fig:EventRate} by the grey dashed lines.\footnote{We note that there is a factor $\sim 3$ difference in the event rate between our result and Ref.\,\cite{Essig:2018tss}.} The grey shaded regions represent variation in the event rate for high and low ionization models of $n_e$\,\cite{Essig:2018tss}. Since there is a difference between three ionization models in the low $n_e$/energy bins, hence we observe a large change in the differential event rates at those bins.  The discovery limits for low and high ionization models is given in appendix\,\ref{app:neUn}. For one electron threshold, the impact of the ionization model uncertainty leads to less than a factor of $3$ change in the discovery limits.

\subsection{Statistical methodology}
\label{subsec:stameth}
In this section, we discuss the statistical procedure to obtain the discovery limit for DM-electron scattering in the presence of substructures for liquid xenon experiments. We have employed the profile likelihood ratio test \cite{Cowan:2010js} with  $\bar{\sigma}_{e}$ and substructure fraction ($\delta$) as the signal parameters of interest. In the following, we briefly discuss this procedure.

The binned likelihood for the background and signal model ($\mathcal{M}$), is given by
\begin{equation}
\mathcal{L}(\mathcal{D}(\theta)|\mathcal{M}(\theta))=\prod_{i=1}^{n_{\rm bins}}\left( \mathcal{P}(N_{\rm obs}^i|N_{\chi}^i+\sum_{j=1}^{n_{\nu}}n_{\nu}^i(\Phi^j)) \right)\prod_{j=1}^{n_{\nu}}\mathcal{G}(\Phi^j)
\label{eq:llhood}
\end{equation}
Here $\theta = (m_{\chi},\bar{\sigma}_e,\delta,\Phi)$ and $\mathcal{D}(\theta)$ is the Asimov data set. The number of energy bin is represented by $n_{\rm bins}$. The Poisson probability ($\mathcal{P}$) at the $i$-th bin is calculated using observed $N_{\rm obs}^i$ and the expected number of events. The expected number of events is the addition of DM events ($N_{\chi}^i$) and the sum of neutrino events ($n_{\nu}^i$) for all the neutrino components ($n_{\nu}$). The Gaussian function ($\mathcal{G}(\Phi^j)$) takes care of the uncertainty in the neutrino fluxes ($\Phi^j$) with mean values and standard deviation given in \cite{Essig:2018tss, OHare:2016pjy}.

Depending on the choice of the analysis, we vary one of the signal parameters (either $\bar{\sigma}_{e}$ or $\delta$), treating the other one as a nuisance parameter. We treat $\bar{\sigma}_{e}$ as the signal parameter for the discovery reach. Therefore, the profile likelihood ratio test statistic, which compares the background-only hypothesis ($\mathcal{M}_0$) with the background and signal model ($\mathcal{M}$), is given by \cite{Cowan:2010js, OHare:2020lva, Buch:2020xyt}
\begin{equation}
q_0 = -2 \, {\rm ln}\left(\frac{\mathcal{L}(\bar{\sigma}_{e}=0,\boldsymbol{\lambda}|\mathcal{M}_0)}{\mathcal{L}\left(\bar{\sigma}_{e},\boldsymbol{\lambda}|\mathcal{M} \right)} \right) \sim \chi_1^2,
\label{eq:q0Dis}
\end{equation}  
where $\boldsymbol{\lambda}$ contains the nuisance parameters, i.e., $\delta$ and $\Phi^j$ in this case. Using Wilks' theorem, it can be shown that the ratio in Eq.\,\eqref{eq:q0Dis} follows a $\chi^2$ distribution with one degree of freedom \cite{Cowan:2010js, Buch:2020xyt}. Thus, the significance of rejecting the background-only hypothesis is given by $\sqrt{q_0}$-$\sigma$. In this paper, we present all the discovery limits at the $90\%$ confidence level (CL). We obtain the discovery limits utilizing Asimov data set which assumes that the number of observed events is same as the expected events. However in a real experimental data-set, this will not be true and in that case one should treat experimental data as the observed events. Then it would be possible to constrain $\bar{\sigma}_e$ assuming a value of substructure fraction.

\begin{figure*}[t]
\begin{center}
	\subfloat[\label{sf:lowexclusion}]{\includegraphics[angle=0.0,width=0.45\textwidth]{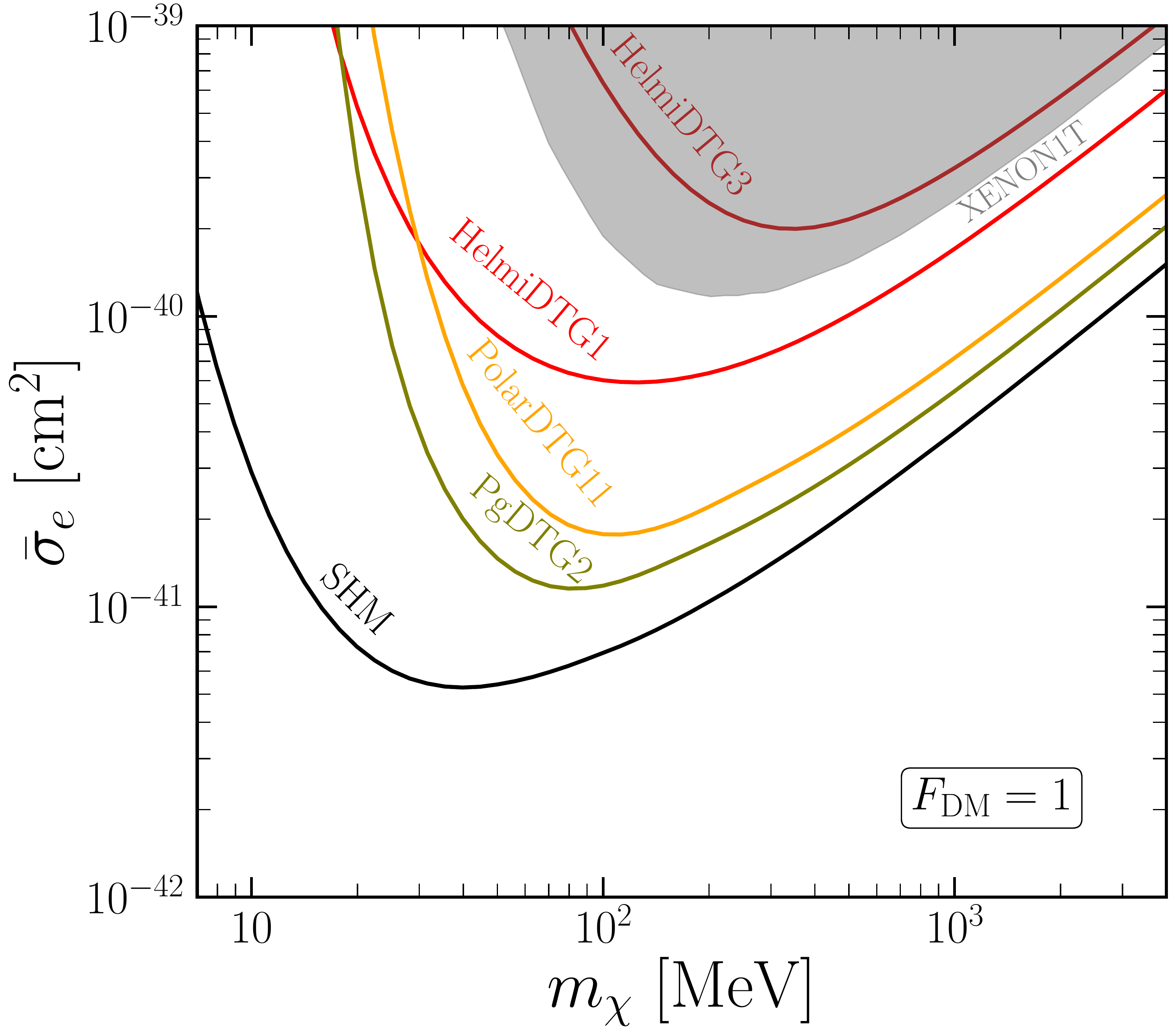}}~~
	\subfloat[\label{sf:highexclusion}]{\includegraphics[angle=0.0,width=0.45\textwidth]{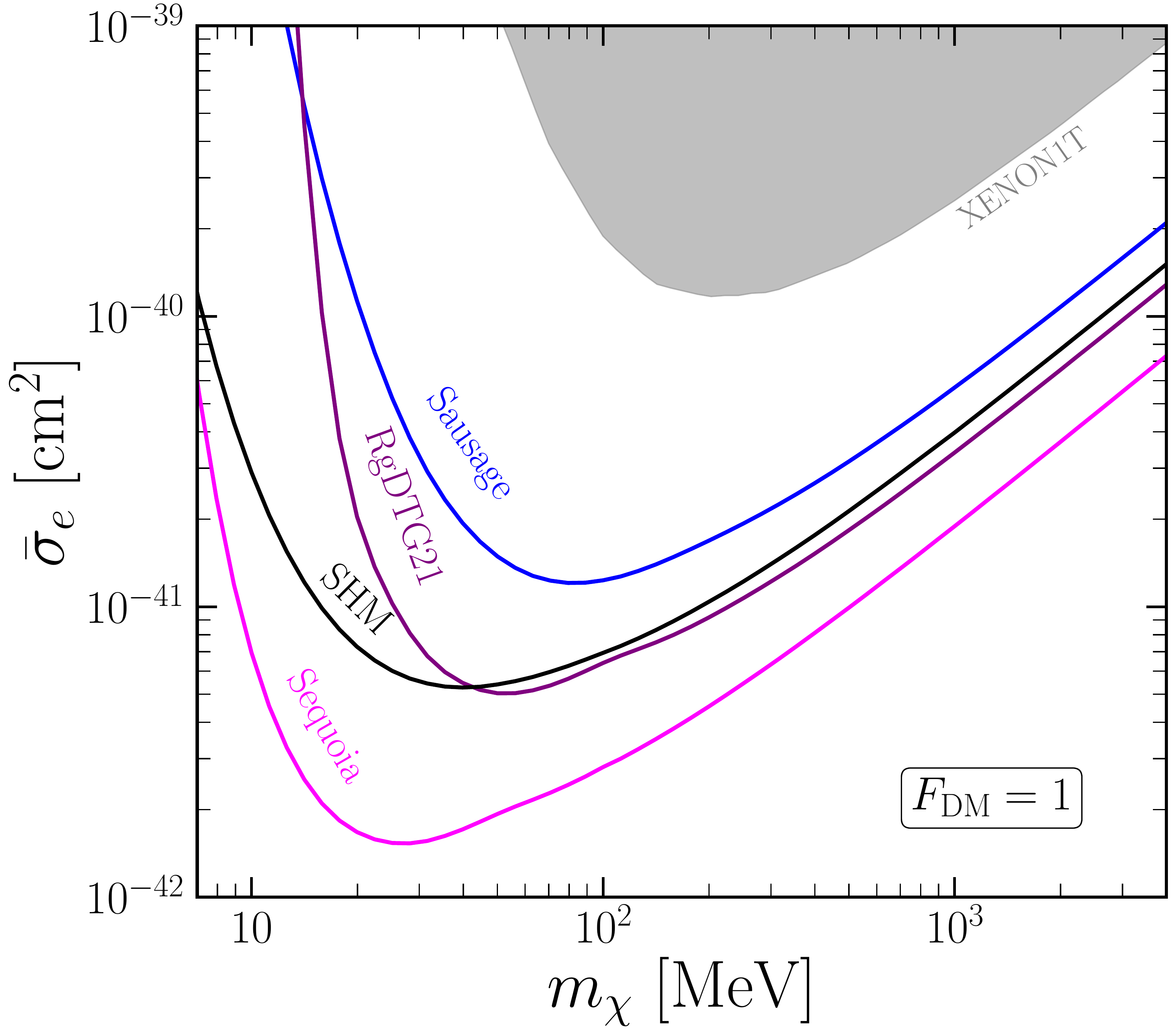}}~~\\	
	\caption{Discovery limits at the $90\%$ CL for different DM substructures considered in this paper, assuming that each astrophysical component accounts for $100\%$ of the local DM density. The discovery limits are obtained using one kg-year exposure, one electron threshold, and $F_{\rm DM}=1.$ }
	\label{fig:Exclusion}
\end{center}	
\end{figure*}

We consider $\delta$ as the signal parameter for the prospective detection of DM substructure fraction. The corresponding profile likelihood ratio test to distinguish two neighboring points $\delta_1$ and   $\delta_2$ can be written as \cite{Buch:2020xyt}
\begin{equation}
q_0 = -2\, {\rm ln}\left(\frac{\mathcal{L}(\delta_2,\boldsymbol{\lambda}|\mathcal{M}_{\delta_1})}{\mathcal{L}\left(\delta_2,\boldsymbol{\lambda}|\mathcal{M}_{\delta_2} \right)} \right) \sim \chi_1^2.
\label{eq:q0focast}
\end{equation}
This profile likelihood ratio is employed to reject the null hypothesis, which is that two neighboring points $\delta_1$ and   $\delta_2$ are indistinguishable at $68\%$ CL. Both for Eqns.\,\eqref{eq:q0Dis} and \eqref{eq:q0focast} we utilized Asimov data set \cite{Cowan:2010js} to obtain the likelihood ratio test. In this scenario, artificial data is generated using the model's parameters (in our case $\mathcal{M}$). Then the expectation is that the number of observed events ($N_{\rm obs}$) should be equal to the number of the expected event ($N_{\rm exp}$). For a sufficiently large number of observations, the value of the profile likelihood ratio test approaches the median value. Compared to the Monte Carlo simulation, the Asimov data set scenario is computationally more economical while acquiring accurate results. For the $68\%$ and $90\%$ CL  limit the required $q_0$'s are $0.99$ and $2.71$ respectively. For a fixed $m_{\chi}$ and $\delta$, the $90\%$ CL discovery limit is obtained by changing $\bar{\sigma}_{e}$ in Eq.\,\eqref{eq:q0Dis} until the required $q_0$ ($2.71$) is achieved. The $68\%$ CL contours in resolving substructure fraction are estimated using Eq.\,\eqref{eq:q0focast}. In this case for a fixed values of $m_{\chi}$, $\sigma_e$, and $\delta_1$, we iterate over $\delta_2$ until the required $q_0$ (= $0.99$) is attained.

\subsection{Results}
\label{subsec:res}
Here we will present the results using the statistical analysis discussed in the previous subsection. The three parameters of interest are DM mass ($m_\chi$), DM-electron cross section ($\bar{\sigma}_e$), and the DM substructure fraction ($\delta$). Given that DM has to be massive, we present our results through two possible choices, keeping one of the other two parameters to a fixed value.   In the first part, the results are presented through the discovery limit, which is depicted in DM mass and DM-electron cross-section plane keeping a fixed DM substructure fraction. In the other case, considering a fixed DM-electron cross-section, we present the forecast of the xenon experiments to resolve the substructure fraction for a few benchmark choices of DM particle masses.
\begin{figure*}[t]
\begin{center}
	\subfloat[\label{sf:lowexclusiondel}]{\includegraphics[angle=0.0,width=0.45\textwidth]{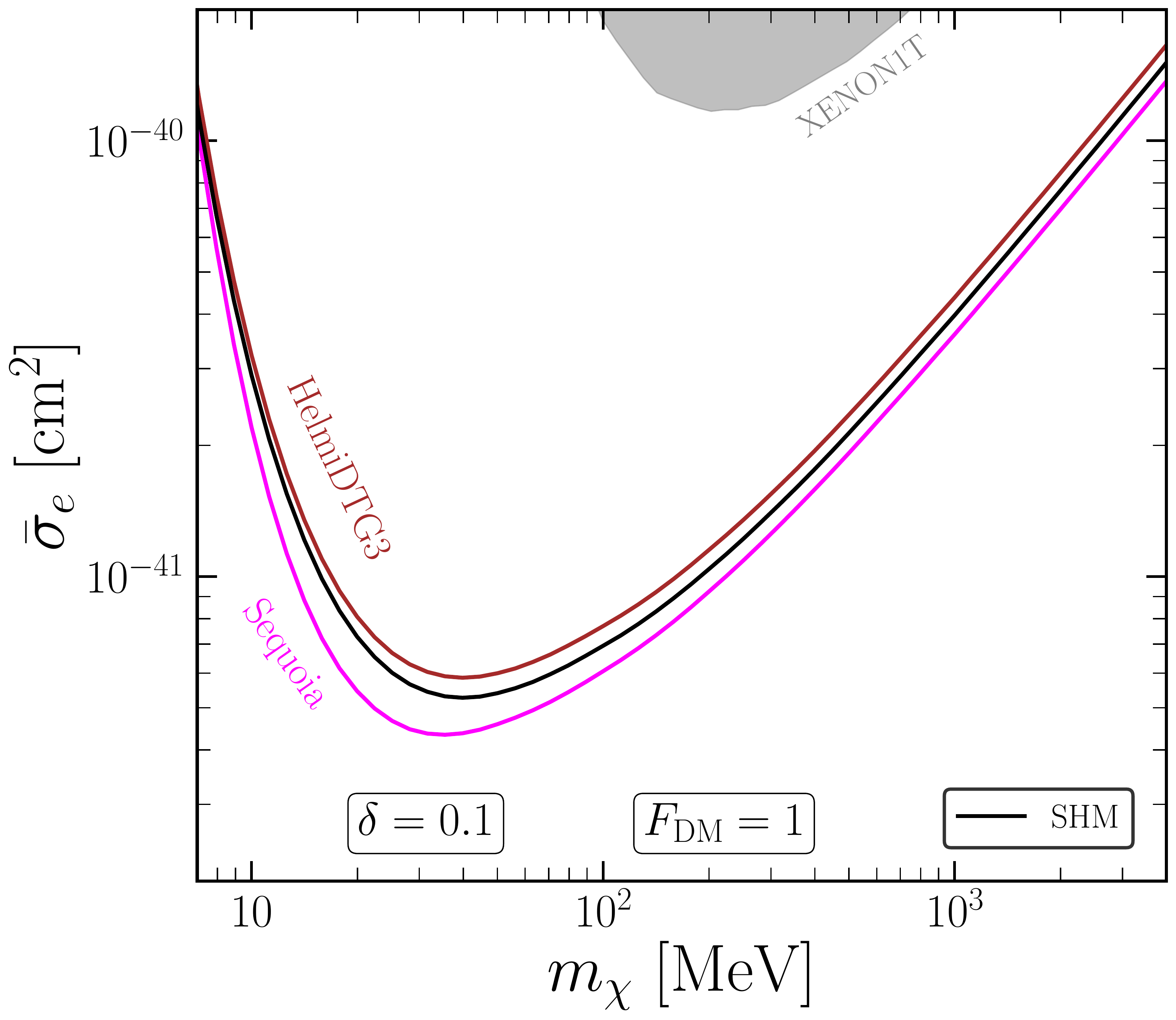}}~~
	\subfloat[\label{sf:highexclusiondel}]{\includegraphics[angle=0.0,width=0.45\textwidth]{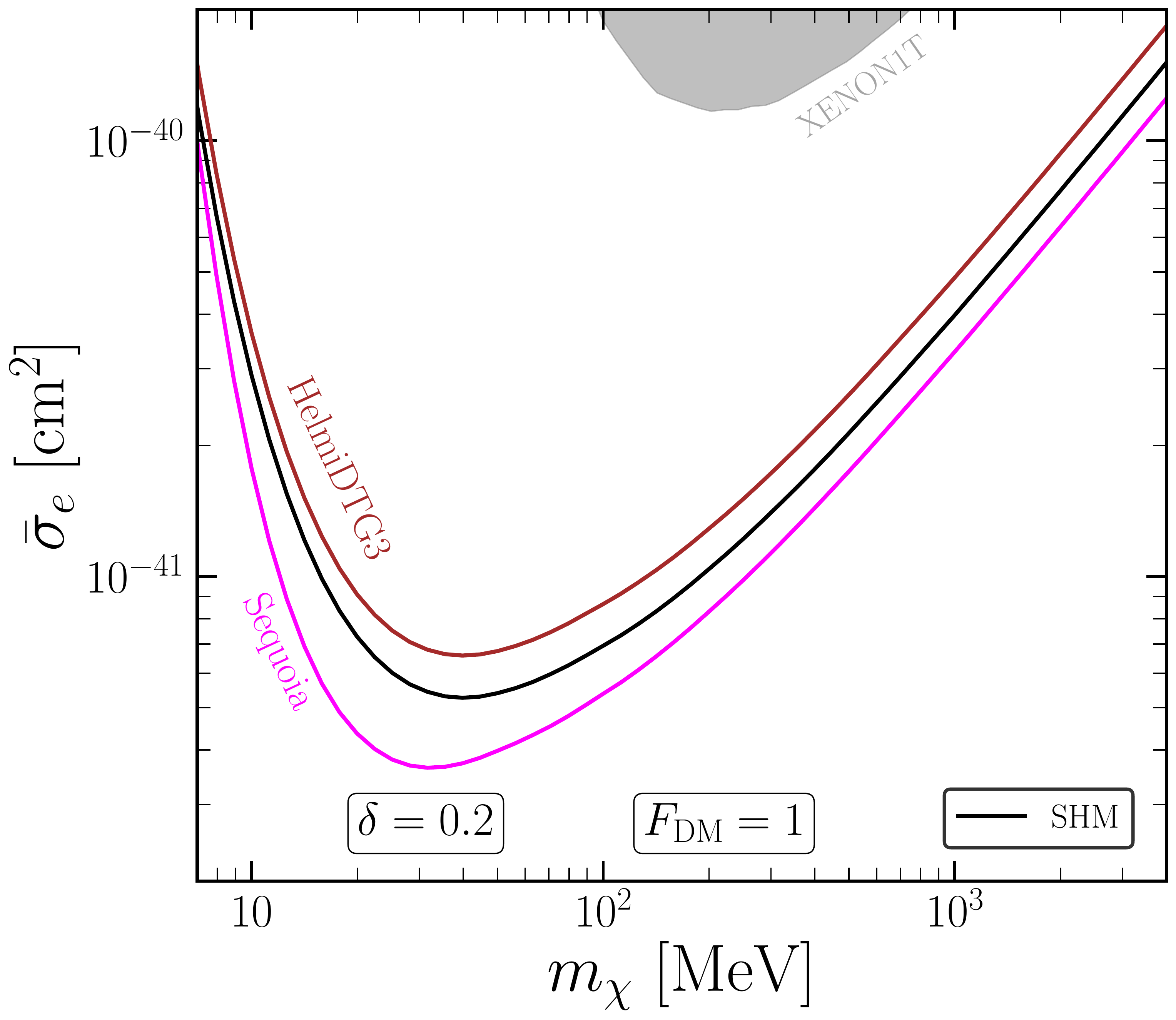}}~~\\	
	\caption{Discovery reach of the DM-electron cross-section at the $90\%$ CL with combined velocity distributions. SHM combined with HelmiDTG3 and Sequoia are shown by the brown and pink solid lines, respectively. The solid black line demonstrates the pure SHM. The discovery limit presented here is for one kg-year exposure, $F_{\rm DM}=1$, and one electron threshold. In the left and right panel we show the results for $\delta=0.1$ and $\delta=0.2$, respectively.}
	\label{fig:Exclusiondel}
\end{center}	
\end{figure*}
In Fig.\,\ref{fig:Exclusion}, we present the sensitivity to DM-electron cross-sections for each of the substructures considered in this paper, assuming that the corresponding substructure constitutes $100\%$ of the local DM density. In Fig.\,\ref{fig:Exclusion}, each line represents the minimum DM-electron cross section required to observe the effect of the corresponding substructure in a liquid xenon detector with $1$ kg-year exposure and one electron threshold. The discovery limits for two and three electron thresholds are given in appendix\,\ref{app:ne2and3}. The different discovery limits for different substructures are the implication of non-identical most probable speed. The tail of the DM velocity distribution will be more populous for the substructure having a relatively larger most probable speed. Therefore a sizable number of DM particles will be available to interact with the target electrons. This leads to a larger event rate, as has been depicted in Fig.\,\ref{fig:EventRate}, where for a fixed DM-electron cross section among the considered DM substructures, we obtain the minimum and the maximum number of events for HelmiDTG3 (lowest most probable speed, see Fig.\,\ref{sf:lowfv}) and Sequoia (highest most probable speed, see Fig.\,\ref{sf:highfv}) respectively. Owing to this, the DM-electron cross-section that can be probed for HelmiDTG3 is the largest, whereas the same for Sequoia is the lowest. The event rates and subsequently the discovery limits lie between HelmiDTG3 and Sequoia for the other considered substructures. The light grey shaded region demonstrates the constraint from the ionization signals in the XENON1T experiment \cite{XENON:2019gfn}, which is the most stringent current DD constraint for the parameter space shown in the plot. For reference, we have also shown the discovery limit for the SHM with the solid black line.

In reality, these substructures may not contribute $100\%$ to the local DM density. Therefore, we choose two benchmark values of $\delta$, namely $\delta=0.1$ and $\delta=0.2$ (shown in Fig.\,\ref{fig:Exclusiondel}).  Further, as mentioned above, we have only considered two substructures, HelmiDTG3 and Sequoia, which lie at two extreme ends. SHM constitutes the rest of the local DM density for the combined DM distribution. If the discovery limit for a particular substructure (with $\delta=1$) is larger compared to SHM, then the same for the combined DM distribution will lie above the SHM limit. This effect would be more pronounced upon increasing $\delta$. In Fig.\,\ref{fig:Exclusiondel}, the combined discovery limit for HelmiDTG3 and Sequoia is displayed by brown and purple lines, respectively. Notably, brown and purple lines lie above and below the SHM scenario. Upon increasing the $\delta$, we observe more deviation from SHM. Importantly, it is still possible to see the effect of these substructures in liquid xenon experiments with this kind of realistic choice of $\delta$.
\begin{figure*}[t]
\begin{center}
	\subfloat[\label{sf:Helmicontour}]{\includegraphics[angle=0.0,width=0.45\textwidth]{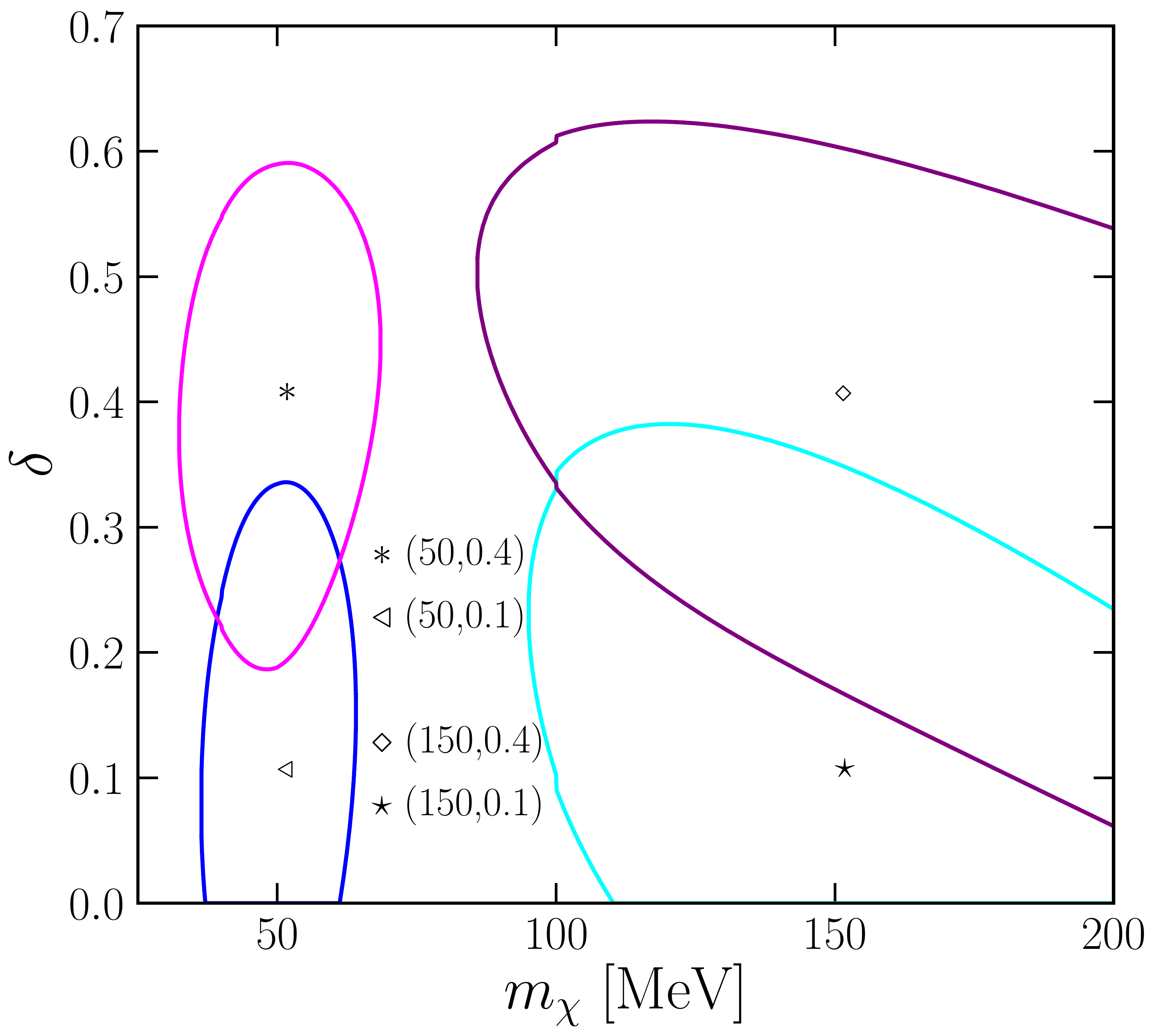}}~~
	\subfloat[\label{sf:Sequoiacontour}]{\includegraphics[angle=0.0,width=0.45\textwidth]{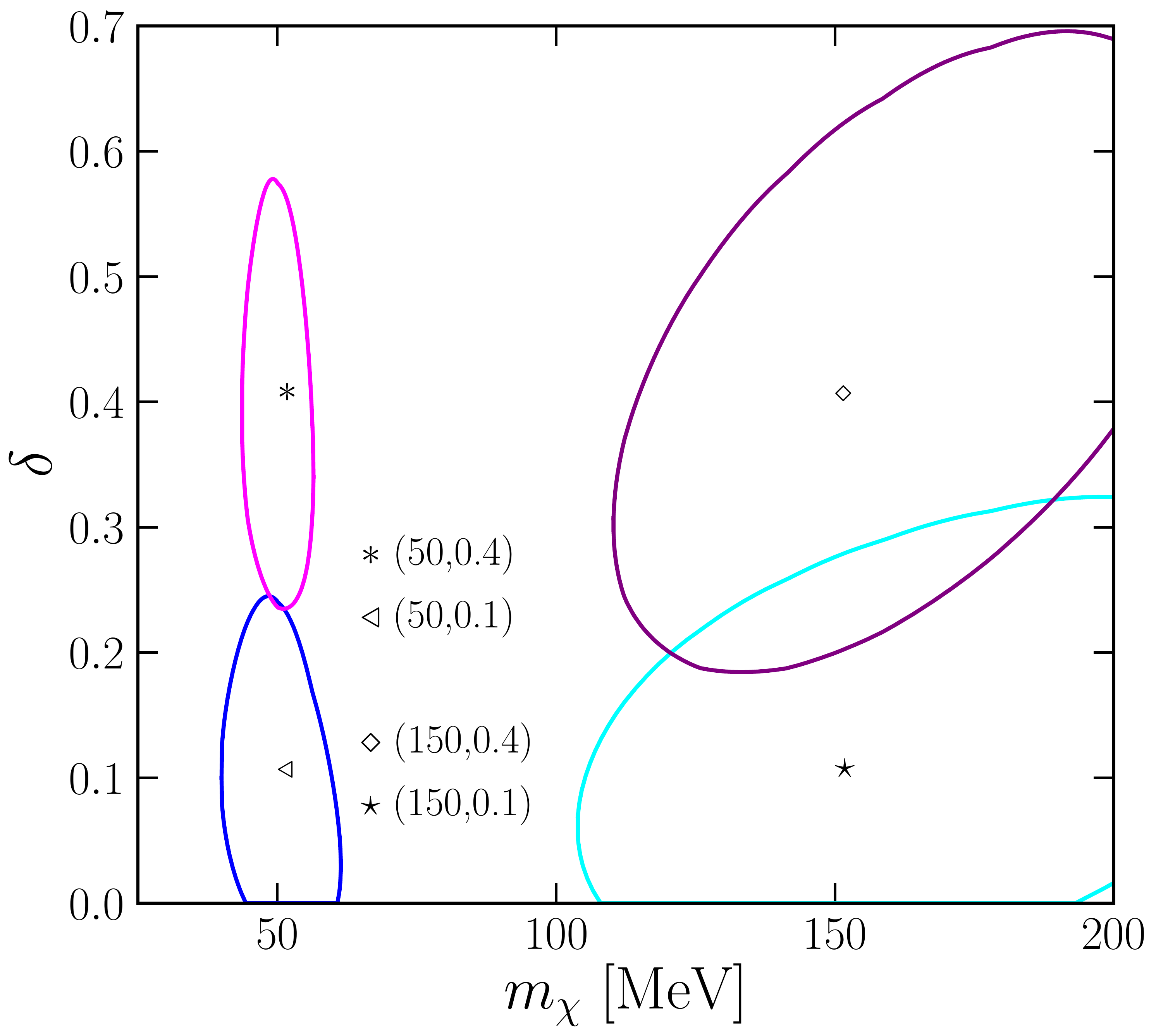}}~~\\	
	\caption{$68\%$ CL contours in $m_{\chi}-\delta$ plane, representing forecast in resolving DM substructure fraction for a few benchmark points. We have assumed 1 kg-year exposure, one electron threshold, $\bar{\sigma}_e=10^{-40}$cm$^2$, and $F_{\rm DM}=1$. In the left and right panels, we show the results for HelmiDTG3 and Sequoia, respectively. }
	\label{fig:contour}
\end{center}	
\end{figure*}
Next, we turn into the discussion of resolving substructure fractions in liquid xenon experiments. Again, we have restricted ourselves to HelmiDTG3 and Sequoia among the considered substructures as these two reside in the extreme ends. The sensitivity in resolving DM substructure at $68\%$ CL is displayed in Fig.\,\ref{fig:contour} for 1 kg-year exposure, one electron threshold, and $\bar{\sigma}_e=10^{-40}\, {\rm cm}^2$, with a few benchmark points. Generically, we observe a better resolution for low DM mass. Comparing Figs.\,\ref{sf:Helmicontour} and \ref{sf:Sequoiacontour} one can see that we will determine the substructure fraction more accurately for Sequoia compared to HelmiDTG3. This is due to Sequoia's large most probable velocity, which leads to a substantial number of DM-electron scattering events. Generically, it is possible to measure the substructure fraction more accurately, which is moving with a higher most probable speed. For $\delta=0.1$, with the considered exposure, threshold, and $\bar{\sigma}_e$, it is difficult to conclude whether the substructure is contributing to local DM density. Interestingly, for DM mass $\sim 50$ MeV, and $\delta = 0.4$, xenon target electron scattering experiments can resolve the substructure fraction with $\sim 50\%$ accuracy. Moreover, the structures of the contours can be understood from Eq.\,\eqref{eq:vmin} and from Fig.\,\ref{fig:eta}. Both for the lower and higher DM masses, the inclination of the contours is reversed as we compare HelmiDTG3 with Sequoia. For low DM masses, $v_{\rm min}$ is larger (for fixed $q$ and $E_e$ from Eq.\,\eqref{eq:vmin}), therefore it is the tail of the distribution which is contributing to $\eta^i$. Thus for HelmiDTG3 with low mass DM, an increment in $\delta$ will reduce the combined value of $\eta$. This reduction could be compensated by increasing DM mass for a fixed number of observed events. This results in a slightly tilted contour towards a higher DM mass. Whereas for higher DM mass (thus smaller $v_{\rm min}$), the maximum value of $\eta$ determines the orientation of contours. For Sequoia, the maximum value of $\eta^i$ is less than that of  HelmiDTG3. Hence increasing $\delta$ for the former will reduce combined $\eta$, which can be elevated by reducing $v_{\rm min}$, i.e., by increasing DM mass.

We have not discussed a distinctive feature of the DM DD signal: annual modulation\,\cite{Lee:2015qva}, where the signal event rates vary with the time of the year in a specified manner. Due to the rotation of the Sun around the MW, there will be DM wind in the Solar rest frame. Due to the Earth's rotational motion around the Sun, the event rate will vary with time. For the SHM, the event rate will be larger (smaller) when the Sun and the Earth travel in opposite (same) directions, respectively. Due to this distinctive feature, which the background cannot mimic, annual modulation events are expected to be less dependent on the background reductions and identifications.

Unlike non-modulation case here we take into account variation of $\mathbf{v_E}$ over time. The main task for the modulation discovery limit would be to evaluate the event rate against both time and energy ($N_{\rm tim}$).  For a particular energy bin, we obtain modulation events ($N_{\rm mod}$), by subtracting each time bin events from average time bin events ($N_{\rm avg}$). We do the same exercise for all the energy bins. The corresponding likelihood can be obtained by taking the difference of their individual Poisson distributions, referred to as the Skellam distribution\,\cite{Skellam1946TheFD}
\begin{eqnarray}
\label{eq:skellam}
\mathcal{L}(\mathcal{D}(\theta)|\mathcal{M}(\theta))
 & \equiv & \prod_{j=1}^{n_{\rm t}} \prod_{i=1}^{n_{\rm bins}}e^{- \left(N_{\rm tim}(i,j)+ N_{\rm avg}(i)\right)} \left( \frac{N_{\rm tim}(i,j)}{N_{\rm avg}(i)}\right)^{N_{\rm mod}(i,j)/2} \nonumber \\ 
&& I_{N_{\rm mod}(i,j)}\left(2\sqrt{N_{\rm mod}(i,j) N_{\rm avg}(i)}\right),
\end{eqnarray}
where, $i$ and $j$ represent each time and energy bin and $n_{\rm t}$ is the total number of time bins. The modified Bessel function of the first kind is  denoted by  $I_{N_{\rm mod}}$. We utilized Eq.\,\eqref{eq:skellam} to obtain test statistics (given in Eq.\,\eqref{eq:q0Dis}) and subsequently the discovery limit. Following this prescription \cite{Buch:2020xyt}, we find that the modulation discovery limit is weaker than the non-modulation counterpart. For example, with SHM or Sequoia, we observed that the modulation discovery reaches are weaker by a factor $\sim 10-100.$  
%
%
%
\section{Conclusions}
\label{sec:conclusion}
The presence of DM in the Universe is well established. Many attempts have been made to discover the connection between DM and SM states. Among them, DD experiments look for the scattering signatures of DM and visible states. There has been a growing interest in the search for light DM (masses $\lesssim 1$\,GeV)  through DD. Ambient non-relativistic DM having mass in the sub-GeV range can not impart sufficient energy to produce a measurable recoil in the typical nuclear recoil DD experiments. Electron, being a light particle, can be an excellent target in detecting such light DM. Many target materials have been considered to identify electronic excitation by the scattering of ambient DM. DM velocity distribution is an integral part of calculating the event rate or the exclusion limit of the DD experiments. DM is also an intrinsic part of structure formation; the history of galaxy formation influences its velocity distribution. While it is difficult to track the velocity distribution of DM, however, it may be manifested through stellar distribution. Surveys like Gaia, SDSS, LAMOST, etc., have made unprecedented progress mapping these stellar distributions. These data reveal the presence of stellar clumps and substructures. It is highly likely that there is a DM counterpart to these stellar substructures, called DM substructure. This paper investigates the prospects of detecting these substructures in low threshold DM DD experiments through elastic DM-electron scattering. Specifically, we have explored the prospect of xenon targets experiments in deciphering this. Note that compared to semiconductor targets experiments (like SENSEI), the xenon targets experiments have better sensitivity in the DM mass range of $\mathcal{O}(100)\,{\rm MeV}$.

We utilize the results of the LAMOST survey and choose a few benchmark DM substructures. We emphasize that there is no definite proof of the existence of the DM counterpart to the detected stellar substructures. However, it is likely that they exist. If these DM substructures overlap with the Earth's position, then we can observe the imprint of the same in xenon targets experiments through DM-electron scattering. We find that if the substructure constitutes $\gtrsim 10\%$ of the local DM density, then there is a possibility to observe the effect of the substructures in xenon target experiments with the currently allowed DM particle properties. We have also explored the forecast of xenon experiments in resolving the DM substructure fraction. We find that the uncertainty in resolving DM substructure fraction is considerable for higher DM mass compared to lower DM mass. For example, with $m_{\chi}=50\,$MeV, $\bar{\sigma}_{e}=10^{-40}{\,\rm cm}^2$, and one electron threshold in xenon experiments, we can resolve the substructure  fraction to $\sim 50\%$ accuracy provided $\delta \sim 0.4.$ The discovery limit and resolving DM substructure fraction are mainly regulated by the most probable velocity of the corresponding velocity distribution. Given this correlation between DD rates and DM velocity distributions, a more detailed understanding of DM substructure is required. High-resolution cosmological simulations and near-future observations will play a crucial role in understanding this. We encourage the experimentalists to continue their excellent work in improving their detector sensitivity so that we are sensitive to such a signal. Our work shows that by pursuing this technique, we will be able to know more about the particle physics and astrophysics of DM and maybe even discover it.

\paragraph*{Acknowledgments\,:} We thank Jatan Buch, Ciaran A.\,J.\,O’Hare,  Mukul Sholapurkar, and Tien-Tien Yu for useful correspondence. We thank John F.\,Beacom, Ciaran A.\,J.\,O’Hare, and Tien-Tien Yu for comments on the manuscript. TNM thanks IOE-IISc fellowship program for financial assistance. RL acknowledges financial support from the Infosys foundation (Bangalore), institute start-up funds, and Department of Science and Technology (Govt. of India) for the grant SRG/2022/001125. 

\appendix
\section{Event rate}
\label{app:evrate}
\begin{figure*}[t]
\begin{center}
	\subfloat[\label{sf:lowevent30MeV}]{\includegraphics[angle=0.0,width=0.45\textwidth]{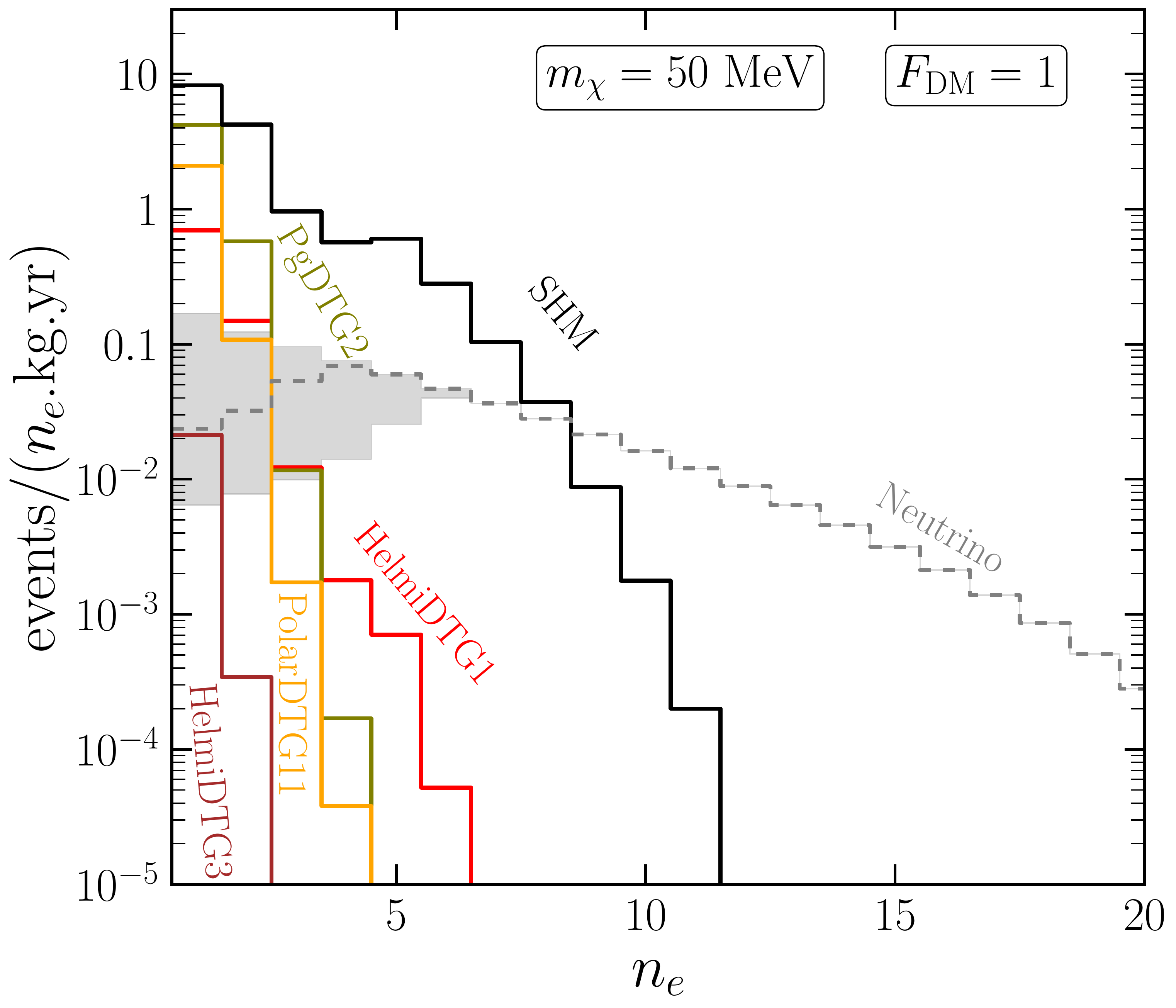}}~~
	\subfloat[\label{sf:highevent30MeV}]{\includegraphics[angle=0.0,width=0.45\textwidth]{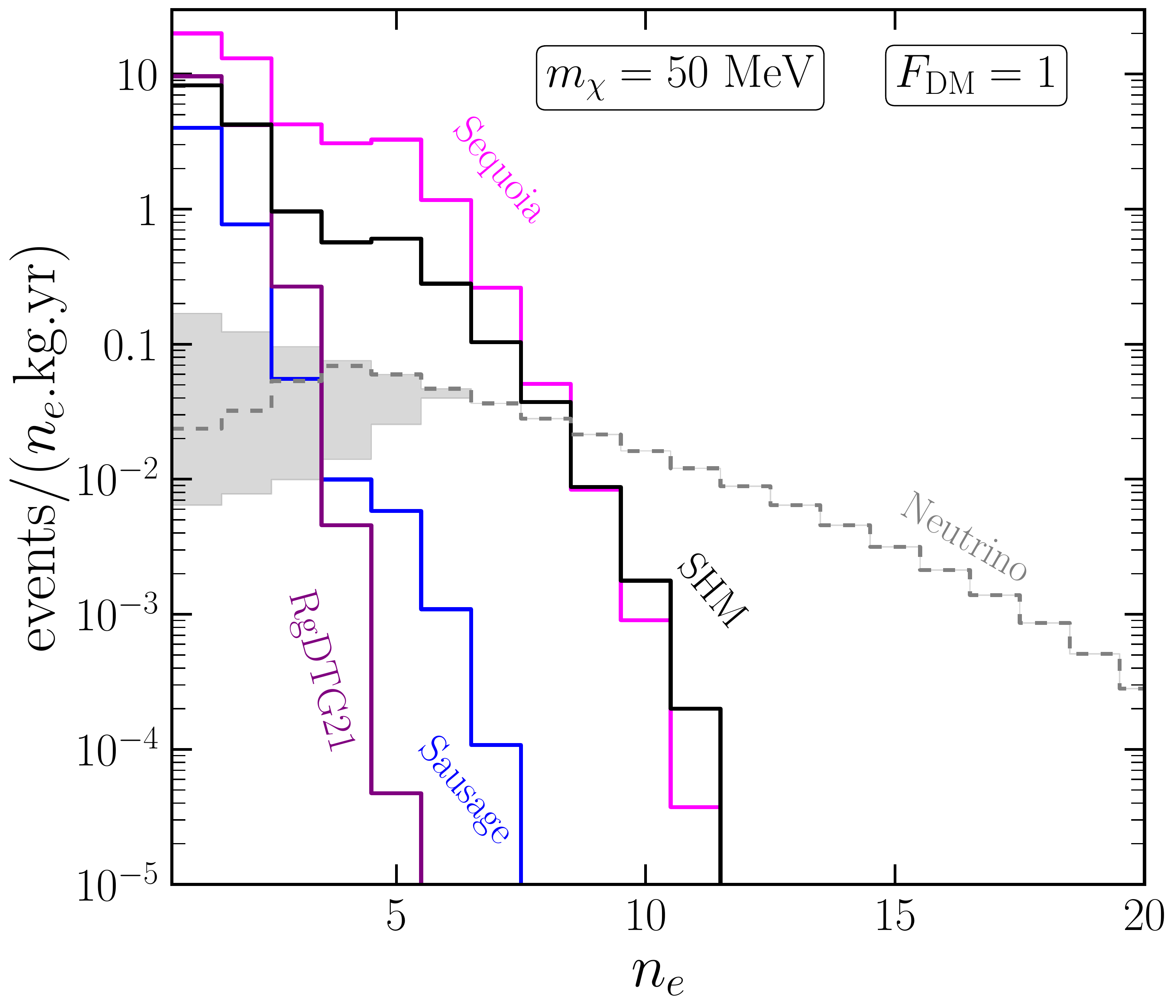}}~~\\	
	\caption{Differential event rate as a function of number of electron $n_e$ for $m_{\chi}=50$\,MeV. Other relevant details are same as Fig.\,\ref{fig:EventRate}.}
\label{fig:EventRate50MeV}
\end{center}	
\end{figure*}

In this appendix we provide event rate for DM mass $50$ MeV. This is displayed in Fig.\,\ref{fig:EventRate50MeV}. 
\section{Discovery limits for two and three electron threshold}
\label{app:ne2and3}
\begin{figure*}[t]
\begin{center}
	\subfloat[\label{sf:ne20p1}]{\includegraphics[angle=0.0,width=0.45\textwidth]{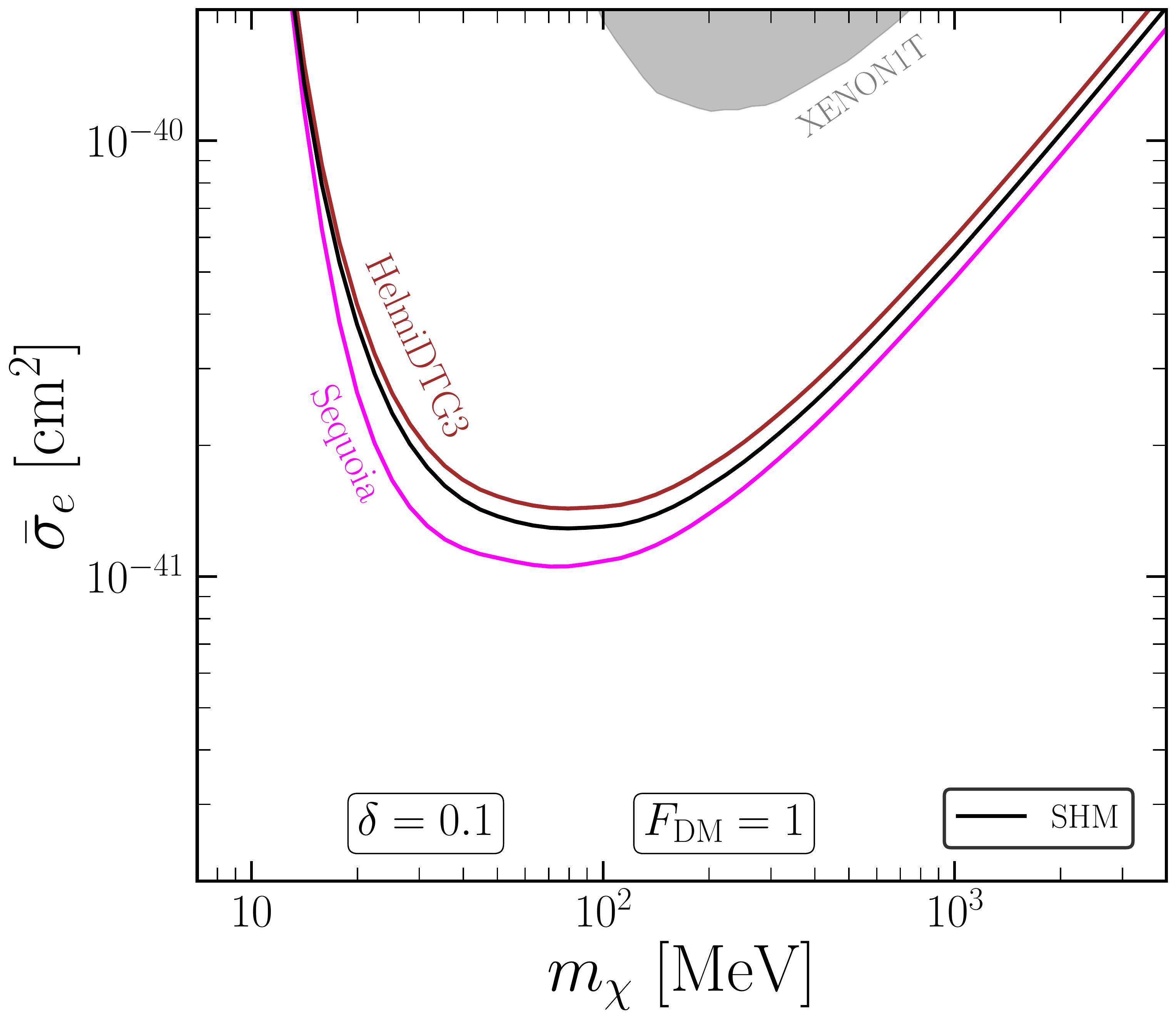}}~~
	\subfloat[\label{sf:ne30p1}]{\includegraphics[angle=0.0,width=0.45\textwidth]{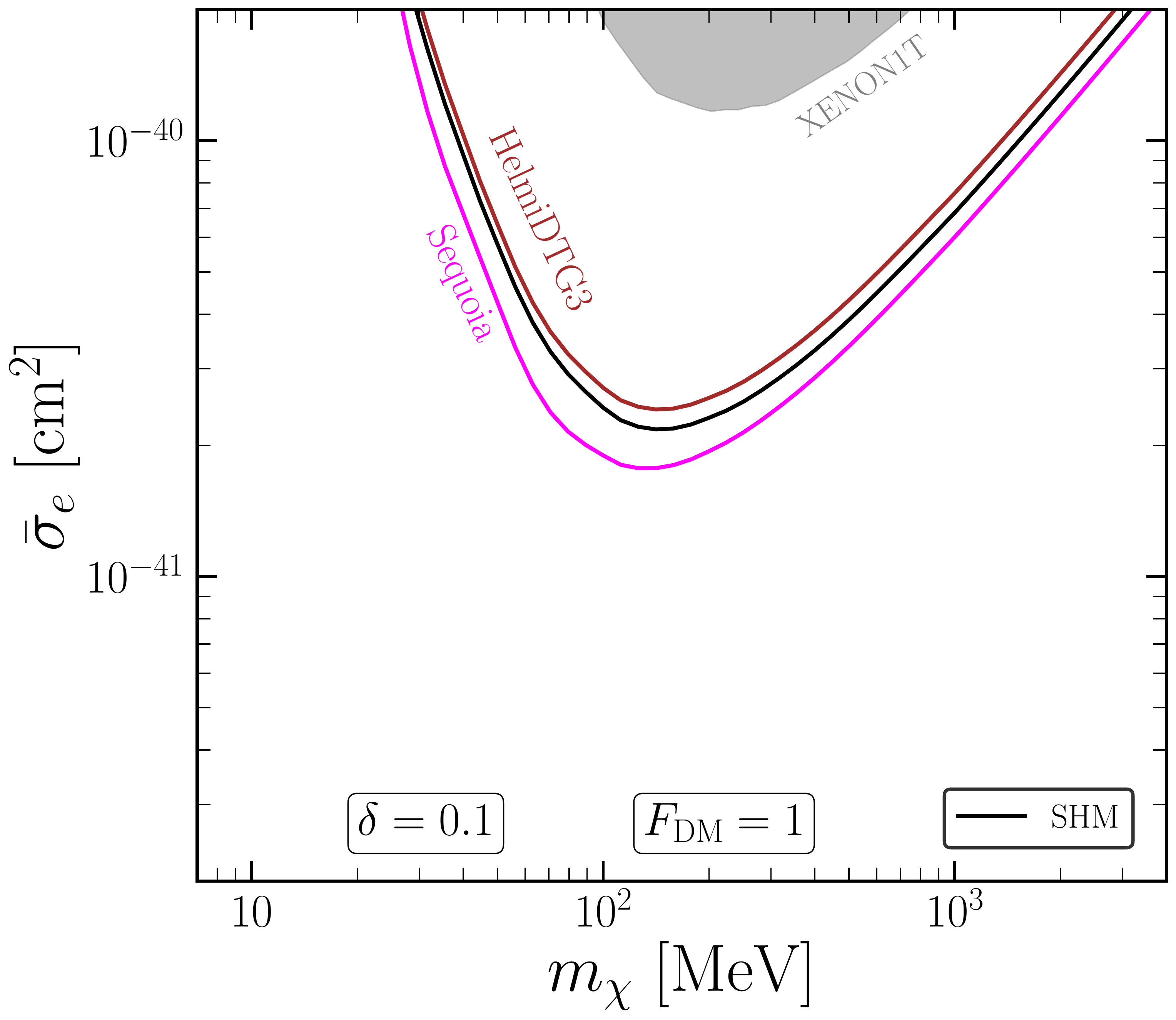}}~~\\	
	\caption{Discovery limits at the $90\%$ CL for the considered DM distribution with $F_{\rm DM}=1$, one kg-year exposure. Each of the considered substructures contributes $10\%$ to the DM density. (a) For two electron threshold. (b) For three electron threshold.}
	\label{fig:ne2and3}
\end{center}	
\end{figure*}

Throughout the main text, we have considered the reach of the xenon experiments for one kg-year exposure and one electron threshold with $F_{\rm DM}=1$. Here we present the discovery limit with two and three electron thresholds for $\delta=0.1$. The results are depicted in Figs.\,\ref{sf:ne20p1} and \ref{sf:ne30p1}. With higher thresholds, the expected event numbers decrease; thus, the required cross-section to see the possible effect of the substructure increases. Further, the lowest possible DM mass that can be probed also increases.

\begin{figure*}[t]
\begin{center}
	\subfloat[\label{sf:neUn1}]{\includegraphics[angle=0.0,width=0.45\textwidth]{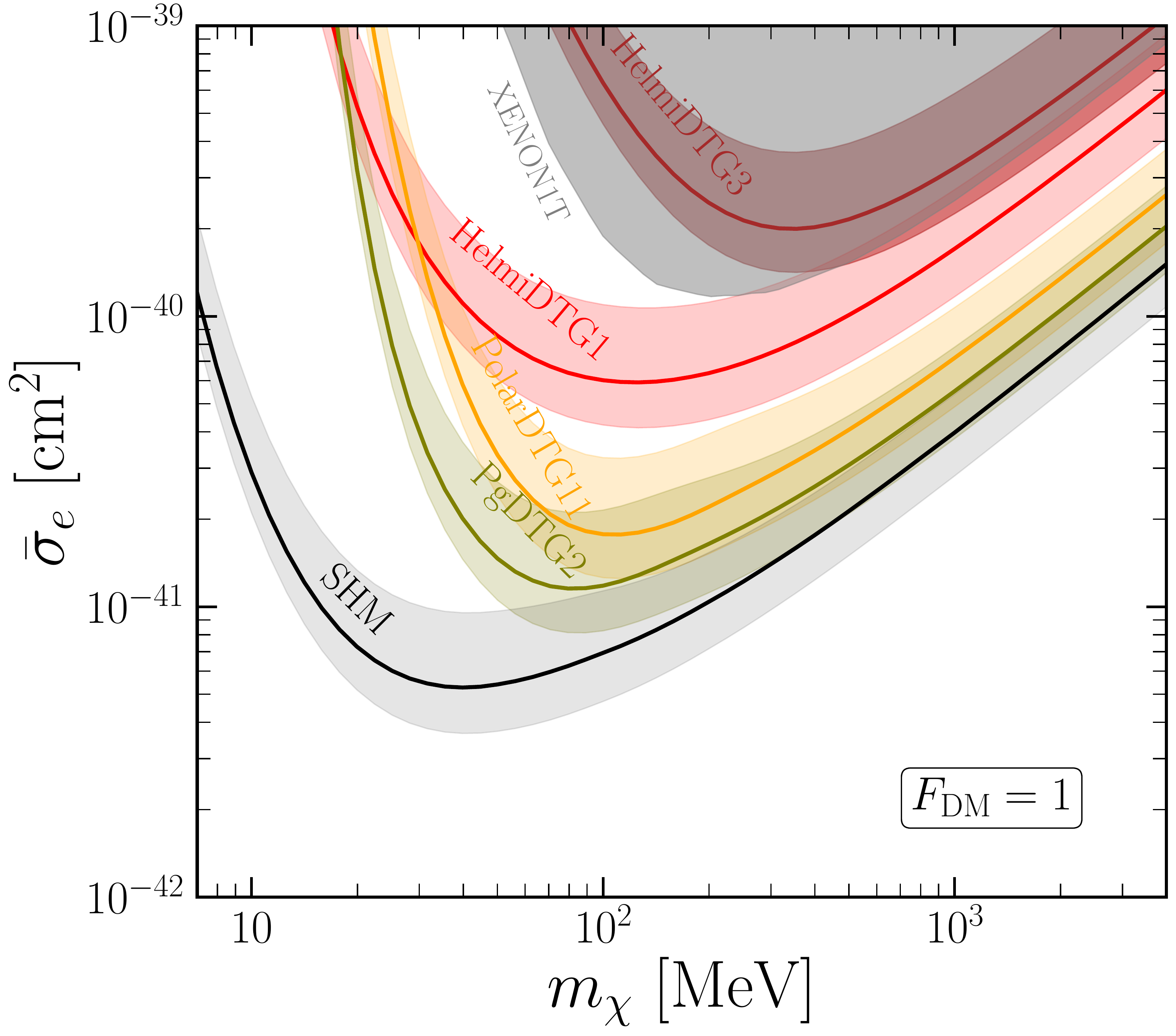}}~~
	\subfloat[\label{sf:neUn2}]{\includegraphics[angle=0.0,width=0.45\textwidth]{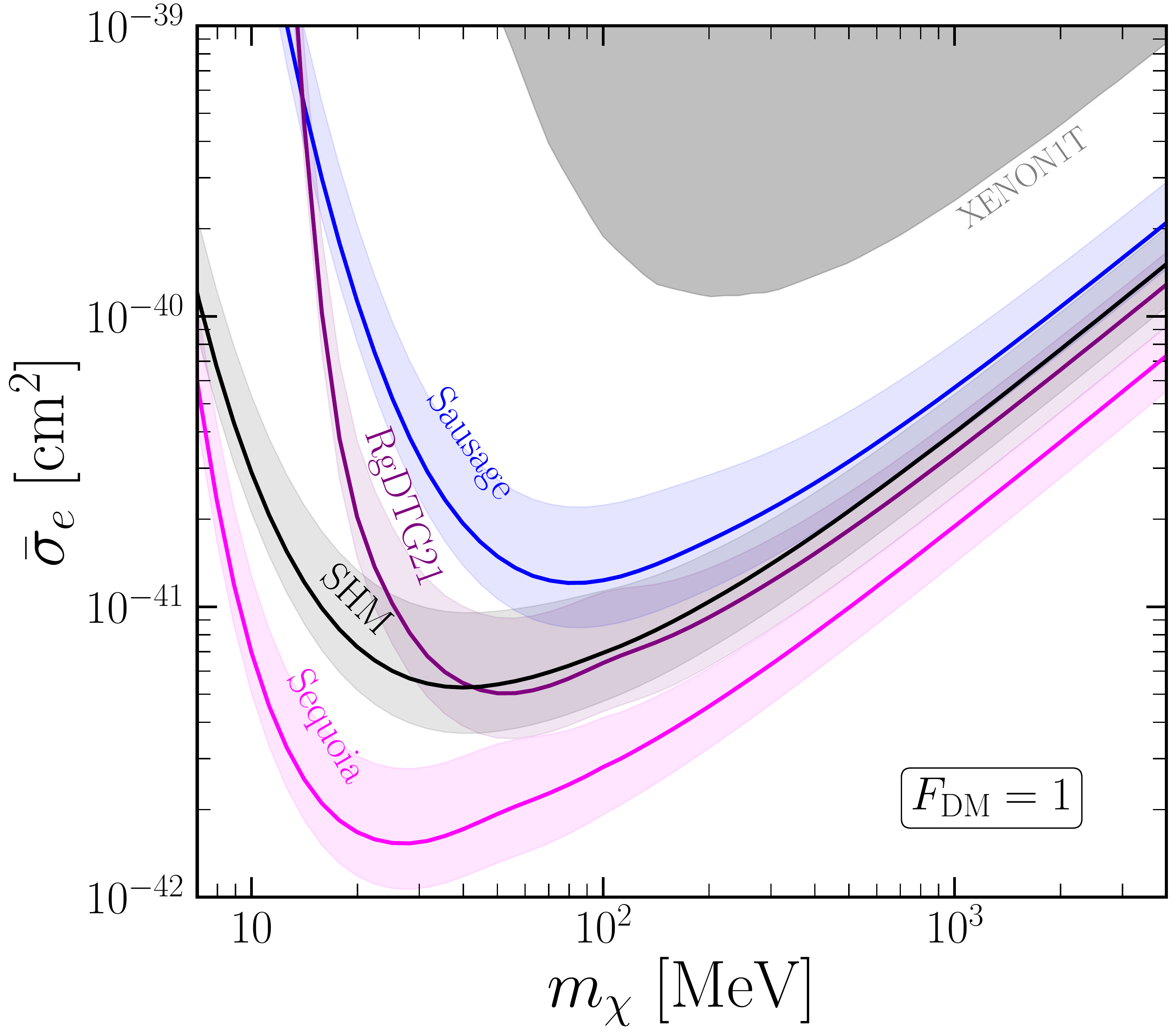}}~~\\	
	\caption{Variation in the discovery limits due to high and low ionization efficiency models for $n_e$\,\cite{Essig:2018tss} assuming one electron threshold. For each of the substructures the corresponding light shaded bands represent the uncertainties that may arise from different ionization models of $n_e$. Each of the substructures contribute to $100\%$ to local DM density.}
	\label{fig:neUn}
\end{center}	
\end{figure*}

\begin{figure*}[]
\begin{center}
	\subfloat[\label{sf:lowDLqm2}]{\includegraphics[angle=0.0,width=0.45\textwidth]{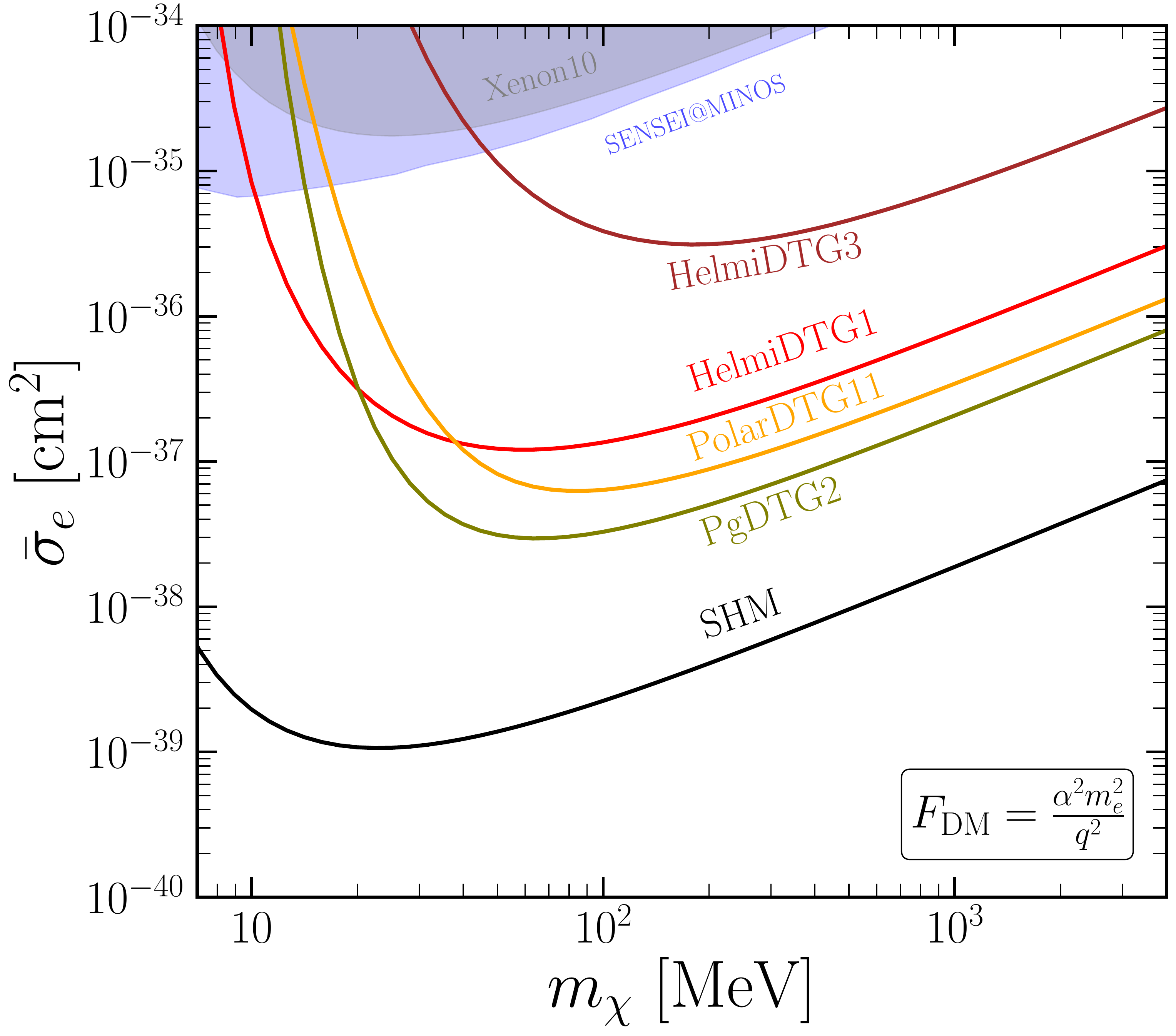}}~~
	\subfloat[\label{sf:highDLqm2}]{\includegraphics[angle=0.0,width=0.45\textwidth]{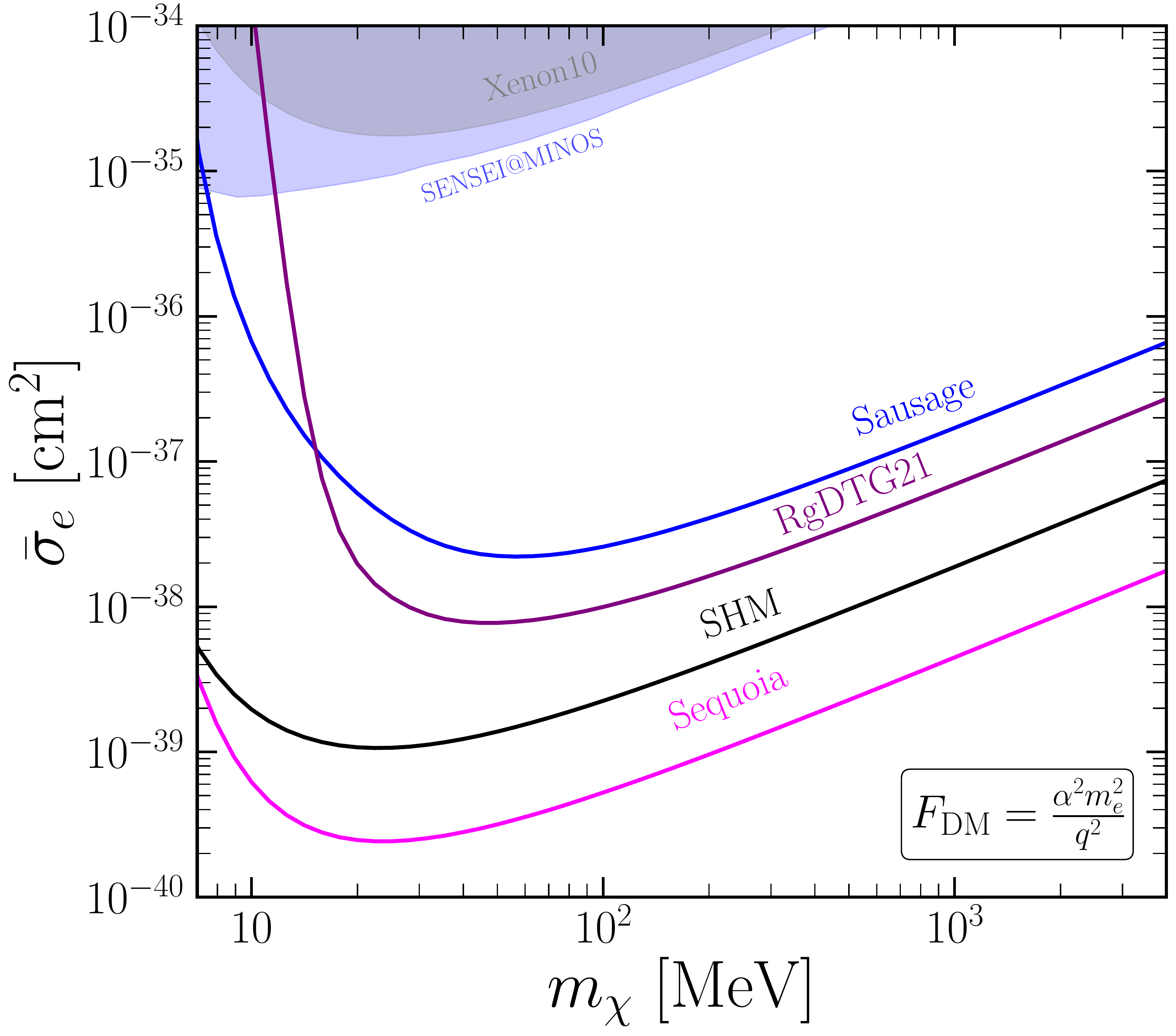}}~~\\	
	\caption{Discovery limits at the $90\%$ CL for the considered DM distribution with $F_{\rm DM}=\alpha^2 m_e^2/q^2$. We have assumed one kg-year exposure and one electron threshold to obtain the discovery limit. The Xenon10\,\cite{Essig:2017kqs} and SENSEI@MINOS\, \cite{SENSEI:2020dpa} limit has also been shown by the grey and blue shaded regions.}
	\label{fig:FDmqm2}
\end{center}	
\end{figure*}

\section{Variation in the discovery limits}
\label{app:neUn}
As discussed in Sec.\,\ref{subsec:nubag}, background event rate from neutrino may change depending on the ionization model. In this appendix, we present the discovery limit for high and low ionization efficiencies models for $n_e$\,\cite{Essig:2018tss}. We display the result in Fig.\,\ref{fig:neUn}. For each of the substructures, solid lines represent discovery limits for fiducial ionization model and shaded bands show the corresponding uncertainties associated with the ionization models.

\section{Momentum dependent DM-electron scattering}
\label{app:FDMqm2}
In this appendix, we present the discovery limits of the momentum-dependent DM-electron scattering, namely $F_{\rm DM}=\alpha^2 m_e^2/q^2$ for the considered DM substructures. In this case, we also observe a similar tendency, except that the minimum required DM-electron cross section for the discovery of the substructures is larger than the same of $F_{\rm DM}=1$. This is displayed in Fig.\,\ref{fig:FDmqm2}.

\bibliographystyle{JHEP}
\bibliography{ref.bib}

\end{document}